\documentclass[english,aps,prl,reprint,twocolumn,nofootinbib,utf8,superscriptaddress,floatfix]{revtex4-2}

\usepackage[T1]{fontenc}
\usepackage{CJKutf8}
\usepackage{color}
\usepackage{mathtools}
\usepackage{amsmath}
\usepackage{amssymb}
\usepackage{graphicx}
\usepackage{wasysym}
\usepackage{indentfirst}
\usepackage{authoraftertitle}
\usepackage{amsmath}
\usepackage{graphicx}
\usepackage{titlesec}

\usepackage[varg]{txfonts}
\usepackage{newtxtext}
\usepackage{bm}

\usepackage[dvipsnames]{xcolor}
\definecolor{pdarkblue}{rgb}{0.1797, 0.1875, 0.5703}
\usepackage{hyperref}
\hypersetup{
 colorlinks = true,
 citecolor = pdarkblue,
 linkcolor = pdarkblue,
 urlcolor = pdarkblue,
}

\usepackage[all]{hypcap}

\AtBeginDocument{
\fontsize{10.5pt}{12.36pt}\selectfont
\fontdimen2\font = 2.9pt 
}

\newlength{\sectionsep}
\renewcommand{\paragraph}[1]{\vspace{\sectionsep}\textit{#1}.---\ignorespaces}
\allowdisplaybreaks

\newcommand{\kB}{k_{\mathrm{B}}}
\newcommand{\T}[1]{T_{\mathrm{#1}}}
\newcommand{\intX}[1]{\int_{\mathcal{X}}\!d\bm{#1}}
\newcommand{\intXX}[2]{\int_{\mathcal{X}\times\mathcal{X}}\!\!\!d\bm{#1}d\bm{#2}}
\newcommand{\Bullet}{\hspace{0.1em}\mathord{\cdot}\hspace{0.1em}}
\newcommand{\doubleBullet}{\hspace{0.1em}\mathord{\cdot}\hspace{0.1em},\mathord{\cdot}\hspace{0.1em}}
\newcommand{\Star}{\scalebox{0.8}{$\mathord{\star}$}}
\newcommand{\ave}[4]{\bar{#1}_{#4}^{#2}(#3)}
\newcommand{\avet}[3]{\bar{#1}_{\,t}^{#2}(#3)}
\newcommand{\aveTt}[3]{\langle\bar{#1}_{\,t}^{#2}\rangle_{#3}}
\newcommand{\aveT}[4]{\langle\bar{#1}_{#4}^{#2}\rangle_{#3}}
\newcommand{\qed}{\hfill$\blacksquare$\vspace{\sectionsep}}
\newcommand{\largestar}{{\mathchoice{\textstyle{*}}{\textstyle{*}}{\scriptstyle{*}}{\scriptstyle{*}}}} 
\newcommand{\piT}[1]{\pi\text{\raisebox{-0.15em}{${}_{#1}$}}}

\begin{document}

\title{\texorpdfstring{Microscopic theory of Mpemba effects and\\ 
a no-Mpemba theorem for monotone many-body systems}
{Microscopic theory of Mpemba effects and a no-Mpemba theorem for monotone many-body systems}}

\author{Naruo Ohga}
\email{naruo.ohga@ubi.s.u-tokyo.ac.jp}
\affiliation{Department of Physics, Graduate School of Science, The University of Tokyo, 7-3-1 Hongo, Bunkyo-ku, Tokyo 113-0033, Japan}

\author{Hisao Hayakawa}
\affiliation{Center for Gravitational Physics and Quantum Information, Yukawa Institute for Theoretical Physics, Kyoto University, Kyoto 606-8502, Japan}

\author{Sosuke Ito}
\affiliation{Department of Physics, Graduate School of Science, The University of Tokyo, 7-3-1 Hongo, Bunkyo-ku, Tokyo 113-0033, Japan}
\affiliation{Universal Biology Institute, Graduate School of Science, The University of Tokyo, 7-3-1 Hongo, Bunkyo-ku, Tokyo 113-0033, Japan}

\date{\today}

\begin{abstract}

Mpemba effects (MPEs), where a hotter system cools faster than a colder one, present intriguing anomalies in relaxation processes. 
Despite their universal observation and significant fundamental and practical implications, a comprehensive theoretical understanding based on microscopic properties remains elusive. 
In this Letter, we introduce two universal frameworks for classical systems to address this gap. Firstly, we reveal that MPEs, traditionally defined by macroscopic temperature comparisons, can be understood through microstate comparisons. 
This insight offers a straightforward and universal microscopic perspective on MPEs, relevant for experiments and numerical simulations to identify their microscopic origins.
Secondly, we establish a ``no-Mpemba theorem,'' a rigorous sufficient condition for the absence of MPEs, thereby identifying specific classes of systems devoid of these effects. 
Our findings are exemplified using ferromagnetic Ising models and one-dimensional multiparticle systems, demonstrating the practical applicability of our theoretical advancements.

\end{abstract}

\maketitle

Thermal relaxation processes are omnipresent in nature, daily life, and industry, often revealing intriguing and complex phenomena. Among these, the Mpemba effect (MPE)—where hot water freezes faster than cold water under certain conditions—stands out~\cite{AristotleMetaphysics,Mpemba1969Cool}. 
Recently, similar effects have been identified across a wide range of systems, including many-body particle systems~\cite{Lasanta2017WhenTheHotterCools,Megias2022KineticTheory,Santos2020MpembaEffect,Takada2021Mpemba,Megias2022ThermalVersusEntropic,Torrente2019LargeMpembaLikeEffect,Biswas2020Mpemba,Mompo2021MemoryEffects,Gonzalez2021MpembaLike,Gonzalez2021TimeDependent,Biswas2021MpembaEffect,Biswas2022MpembaEffect,Megias2022MpembaLike,Takada2021HomogeneousCooling,Patron2021StrongNonexponential,Zuk2022TransientDynamics,Patron2023NonEquilibrium,Ghosh2024SimulationsOfMpemba}, 
classical spin systems~\cite{Vadakkayil2021ShouldAHotterParamagnet,Chatterjee2024MpembaEffect,Das2023PerspectivesOnAFewPuzzles,Ghosh2024SimulationsOfMpemba,Yang2020NonMarkovianMpemba,Zhang2022TheoreticalModel,Holtzman2022LandauTheory,BaityJesi2019MpembaEffect,Yang2022MpembaEffect,Teza2023EigenvalueCrossing,Israel2019PRX}, 
one-dimensional colloidal systems~\cite{Lu2017NonequilibriumThermodynamics,Kumar2020ExponentiallyFaster,Kumar2022AnomalousHeating,Chtrite2021TheMetastableMpemba,Walker2022MpembaEffect,Walker2021AnomalousThermal,Deguenther2022AnomalousRelaxation,Biswas2023MpembaEffect1,Biswas2023MpembaEffect2,Schwarzendahl2022AnomalousCooling,Teza2023RelaxationShortcuts,Walker2023OptimalTransport}, 
small discrete-state systems~\cite{Lu2017NonequilibriumThermodynamics,Israel2019PRX,Busiello2021InducingAndOptimizing,Teza2023RelaxationShortcuts,Walker2023OptimalTransport,Biswas2024MpembaEffect}, 
quantum systems~\cite{Nava2019LindbladDissipative,Ares2023EntanglementAsymmetry,Yamashika2024EntanglementAsymmetry,Murciano2024EntanglementAsymmetry,Srivastav2024FamilyOfExactAndInexact,Rylands2024MicroscopicOrigin,Joshi2024ObservingTheQuantum,Caceffo2024EntangledMultiplets,Liu2024SymmetryRestoration,Chatterjee2023QuantumMpemba,Chatterjee2024Multiple,Shapira2024Inverse,Wang2024MpembaEffects,Strachan2024NonMarkovianQuantum,Nava2024MpembaEffects,Moroder2024ThermodynamicsOfTheQuantum,Longhi2024PhotonicMpemba,Chang2024ImaginaryTime}, and various mesoscopic and macroscopic experimental setups~\cite{Greaney2011MpembaLike,Ahn2016ExperimentalVerifications,Chorazewski2023TheCuriousCase,Lv2023EffectsOfMolecularWeight,Dekhtyar2023ObservationOfAPlasma,Greaney2011MpembaLike,Liu2023MpembaEffectInCrystallization}. 
The MPE has emerged as a captivating topic in nonequilibrium statistical physics, driven by both its fundamental significance and potential technological applications. 
Harnessing MPEs could revolutionize the engineering of accelerated relaxation processes~\cite{Pemartin2024ShortcutsOf,Gal2020Precooling,Gonzalez2021SlowGrowth,Ivander2023Hyperacceleration,Carollo2021ExponentiallyAccelerated,Bao2022AcceleratingRelaxation,Kochsiek2022Accelerating,Chittari2023GeometricApproach}, such as rapid material equilibration~\cite{Gonzalez2021SlowGrowth,Ivander2023Hyperacceleration}, enhanced heat engine performance~\cite{Cao2022FastFunctionalization,Lin2022PowerStatistics}, and accelerated quantum computing~\cite{Kochsiek2022Accelerating,Joshi2024ObservingTheQuantum}.

Two primary scenarios for the MPEs have been identified in the literature. 
The first scenario involves nonequilibrium initial conditions that influence the relaxation speed~\cite{Lasanta2017WhenTheHotterCools,BaityJesi2019MpembaEffect,Ares2023EntanglementAsymmetry,Chatterjee2023QuantumMpemba,Carollo2021ExponentiallyAccelerated}. 
The second scenario, which is the focus of this Letter, examines two equilibrium initial states where MPEs emerge from the system's intrinsic properties, such as potential landscapes~\cite{Lu2017NonequilibriumThermodynamics,Israel2019PRX,Kumar2020ExponentiallyFaster,Kumar2022AnomalousHeating,Holtzman2022LandauTheory}. 
These scenarios have been predominantly analyzed through a combination of numerical simulations and simplified theoretical models involving a few macroscopic degrees of freedom~\cite{Santos2024MpembaMeets}. 
For instance, MPEs in particle gases have been explored using a two-variable dynamical system of temperature and kurtosis~\cite{Lasanta2017WhenTheHotterCools,Santos2020MpembaEffect,Takada2021Mpemba,Megias2022ThermalVersusEntropic,Megias2022KineticTheory}. 
In Markovian classical and quantum systems, MPEs have been studied in terms of the slowest eigenmodes of time evolution~\cite{Lu2017NonequilibriumThermodynamics,Israel2019PRX,Teza2023EigenvalueCrossing,Pemartin2024ShortcutsOf,Chatterjee2023QuantumMpemba,Carollo2021ExponentiallyAccelerated}, which typically correspond to macroscopic modes in large systems~\cite{Pemartin2024ShortcutsOf}.

While these theories capture universal macroscopic mechanisms of MPEs, the reflection of the system's microscopic properties in MPEs remains unclear.
The microscopic mechanisms of MPEs in large systems have predominantly been discussed on a case-by-case basis~\cite{BaityJesi2019MpembaEffect,Walker2021AnomalousThermal,Yang2022MpembaEffect,Teza2023EigenvalueCrossing,Rylands2024MicroscopicOrigin,Yamashika2024EntanglementAsymmetry,Chang2024ImaginaryTime}, lacking a universal theoretical framework. 
Establishing such a framework is crucial for harnessing MPEs across various systems.
Furthermore, it has rarely been possible to analytically prove the existence or absence of MPEs based on the system's microscopic dynamics, except in simple solvable models~\cite{Lu2017NonequilibriumThermodynamics,Walker2021AnomalousThermal,Deguenther2022AnomalousRelaxation,Biswas2023MpembaEffect1,Biswas2023MpembaEffect2,Teza2023EigenvalueCrossing,Ivander2023Hyperacceleration}.
An analytical proof could provide a comprehensive understanding of when MPEs occur and when they do not, potentially settling debates on whether certain classes of systems exhibit MPEs. 
For instance, the existence of MPEs in water remains a topic of contention~\cite{Burridge2016Questioning,Burridge2020Observing,Elton2021PathologicalWater}.

In this Letter, we address these issues by presenting two theoretical results that connect MPEs with microscopic properties. 
First, we show that MPEs are universally understood as a consequence of the accumulation of \textit{microstate MPEs}, microscopic phenomena defined by comparing two microstates. 
This universal framework allows us to decompose the MPE into contributions from each microstate pair, facilitating the analysis of the microscopic origins of MPEs in experiments and numerical simulations.
Second, we propose a ``no-Mpemba theorem'' that provides a sufficient condition for the absence of ensemble MPEs in a broad class of Markovian discrete-state many-body systems.
This condition is based on \textit{monotone coupling}~\cite{Liggett2005InteractingParticle,Levin2017MarkovChains}, a mathematical tool for comparing the statistics of two microstates.
The condition can be verified using the system's microscopic description without solving the time evolution, making it applicable to a wide range of many-body systems, including those with quenched randomness.
We apply this theorem to demonstrate the absence of MPEs in the ferromagnetic Ising model under strong magnetic fields and in the exclusion process within a one-dimensional monotone potential.

\paragraph{Setup and definition of MPE}
We consider a general classical system in contact with a thermal bath, undergoing a possibly underdamped stochastic time evolution. We write a microstate of the system as a vector $\bm{x} \equiv (x_{1},x_{2},\dots)$, which may be high-dimensional for many-body systems and takes either discrete or continuous values. We write the set of all possible microstates as $\mathcal{X} \ni \bm{x}$. 
The system at state $\bm{x}$ has energy $\epsilon(\bm{x})$,  which includes interaction energy for many-body systems. 
The Gibbs distribution at temperature $T$ is $\piT T(\bm{x}) \coloneqq \exp\{-[\epsilon(\bm{x})-F_{T}]/\kB T\}$, where $F_{T} = -\kB T\ln\intX {x}\exp[-\epsilon(\bm{x})/\kB T]$, and $\kB$ is the Boltzmann constant. 
Here and hereafter, a probability distribution refers to a probability density function for continuous systems and a probability mass function for discrete systems, and the volume integral $\intX x$ should always be replaced with $\sum_{\bm{x} \in \mathcal{X}}$ for discrete systems.

We put the system in a cold thermal bath at time $t = 0$ and let it relax. A single realization (trajectory) of this cooling process starts with one of the states in $\mathcal{X}$ and continually moves over $\mathcal{X}$ without ever stopping. We write the probability of being in $\bm{x}$ at time $t$ conditional to being in $\bm{y}$ at time 0 as $P_{t}^{\T b}(\bm{x}\vert\bm{y})$, which depends on the bath temperature $\T b$.
The time evolution may be non-Markovian as long as the system holds no memory from $t < 0$, and therefore $P_{t}^{\T b}(\bm{x}\vert\bm{y})$ is determined by the state $\bm y$ at $t=0$.
The conditional distribution should converge to equilibrium, $\lim_{t \to \infty}P_{t}^{\T b}(\bm{x}\vert\bm{y}) = \piT{\T b}(\bm{x})$, from any $\bm{y}$.

We consider two protocols for preparing the initial state at $t = 0$ before cooling the system and measuring an observable (a function of state) $a(\bm{x})$ at time $t$. In the first protocol, we prepare the system in a single fixed state $\bm{y}$. After evolving the system, the average of $a(\bm{x})$ at time $t$ is given by $\avet a{\T b}{\bm{y}} \coloneqq \intX x\,a(\bm{x})P_{t}^{\T b}(\bm{x}\vert\bm{y})$. In the second protocol, we equilibrate the system with a bath of temperature $\T i$ at $t < 0$ so that the initial state $\bm y$ at $t = 0$ distributes according to $\piT{\T i}(\bm{y})$. This second protocol is equivalent to averaging the first protocol over many initial states. Thus, the average of $a(\bm{x})$ at time $t$ is given by $\aveTt a{\T b}{\T i} \coloneqq \intX y\,\avet a{\T b}{\bm{y}}\piT{\T i}(\bm{y})$. 

\begin{figure}
    \centering
    \includegraphics[width=1\columnwidth]{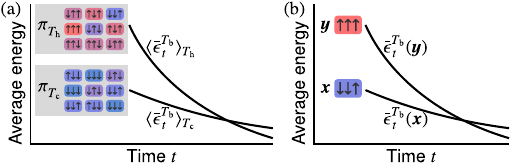}
    \caption{Illustration of concepts.  (a) Ensemble MPE uses two initial Gibbs ensembles (gray boxes), i.e., mixtures of microstates (small rectangles). Color indicates the state energy (red: high; blue: low). (b) Microstate MPE uses two microstates as the initial conditions.}
    \label{fig:concept}
\end{figure}

Let us consider the second protocol with two initial temperatures $\T i = \T c, \T h$ and one bath temperature $\T b$ that satisfy $\T h > \T c > \T b$. We define \textit{the ensemble MPE with initial temperatures $\T c$ and $\T h$ at time $t$} by the condition
\begin{equation}
\aveTt{\epsilon}{\T b}{\T c} > \aveTt{\epsilon}{\T b}{\T h}.
\label{eq:ensemble_def}
\end{equation}
Noting that $\aveT{\epsilon}{\T b}{\T c}{0} < \aveT{\epsilon}{\T b}{\T h}{0}$ at the initial time, this definition can be depicted as the crossing of two curves in Fig.~\ref{fig:concept}(a). 
This comparison of two initial Gibbs distributions is a standard formulation in literature~\cite{Lu2017NonequilibriumThermodynamics,Israel2019PRX,Busiello2021InducingAndOptimizing,Chtrite2021TheMetastableMpemba,Walker2021AnomalousThermal,Kumar2020ExponentiallyFaster,Kumar2022AnomalousHeating,Holtzman2022LandauTheory}, and here we call it the \textit{ensemble MPE} for clarity. The average energy has served as a practical signature of MPEs~\cite{Biswas2020Mpemba,Biswas2022MpembaEffect,BaityJesi2019MpembaEffect,Vadakkayil2021ShouldAHotterParamagnet,Chatterjee2024Multiple,Das2023PerspectivesOnAFewPuzzles,Chatterjee2024MpembaEffect}, especially for many-body systems.

\paragraph{Result 1: Universal microscopic framework of MPEs}
Our first result relates the ensemble MPE to a microscopic phenomenon. To do so, we introduce a microscopic analog of MPE that compares two microstates, which we call the \textit{microstate MPE}\@.
For a bath temperature $\T b$, we say that \textit{the microstate MPE between states $\bm{x}$ and $\bm{y}$ is present at time $t$}
if $\epsilon(\bm{x})-\epsilon(\bm{y})$ and $\avet{\epsilon}{\T b}{\bm{x}}-\avet{\epsilon}{\T b}{\bm{y}}$ have opposite signs, or equivalently,
\begin{equation}
c_{t}^{\T b}(\bm{x},\bm{y}) \coloneqq \bigl[\avet{\epsilon}{\T b}{\bm{x}}-\avet{\epsilon}{\T b}{\bm{y}}\bigr][\epsilon(\bm{x})-\epsilon(\bm{y})] < 0.
\label{eq:microstate_def}
\end{equation}
Since $\ave{\epsilon}{\T b}{\bm{x}}0 = \epsilon(\bm{x})$ and $\ave{\epsilon}{\T b}{\bm{y}}0 = \epsilon(\bm{y})$, this definition can be illustrated as the crossing of two curves in Fig.~\ref{fig:concept}(b).

Result 1 states that, for fixed bath temperature $\T b$ and time $t$, the following two statements are equivalent: \textit{(a) The ensemble MPE occurs at time $t$ for at least one pair of initial temperatures $\T h > \T c$ ($ > \T b$); (b) The average of $c_{t}^{\T b}(\bm{x},\bm{y})$,
\begin{equation}
C_{t}^{\T b}(T_{\largestar}) \coloneqq \intXX xy\,c_{t}^{\T b}(\bm{x},\bm{y})\piT{T_{\largestar}}(\bm{x})\piT{T_{\largestar}}(\bm{y})
\label{eq:main-result1}
\end{equation}
is negative for at least one temperature $T_{\largestar}$ ($ > \T b$)}. Moreover, the temperatures in these statements can be taken to satisfy $\T c < T_{\largestar} < \T h$. More precisely, if (a) holds, we can take $T_{\largestar}$ of (b) between $\T c$ and $\T h$, and if (b) holds, we can take $\T c$ and $\T h$ of (a) so that $\T c < T_{\largestar} < \T h$. We prove this result in Supplemental Material (SM)~\cite{SM} using the fluctuation--dissipation theorem.
\nocite{VandenBroeck2015EnsembleAndTrajectory,Risken1989FokkerPlanck}

This result offers a comprehensive microscopic framework for understanding ensemble MPEs. 
Specifically, the ensemble MPE manifests if and only if the microstate MPEs are prevalent across a significant number of state pairs. 
The microstate MPE is characterized by higher-energy states dissipating energy more rapidly, while lower-energy states retain their energy for extended periods. 
When this behavior is observed across numerous state pairs, the ensemble MPE emerges.
This conceptualization provides an intuitive understanding of ensembles MPE, which is not immediately apparent from the original definition in Eq.~\eqref{eq:ensemble_def}.

Furthermore, Result 1 helps analyze the origin of ensemble MPEs in experiments and numerical simulations. From Eq.~\eqref{eq:main-result1}, we can interpret $c_{t}^{\T b}(\bm{x},\bm{y})\piT{T_{\largestar}}(\bm{x})\piT{T_{\largestar}}(\bm{y})$ as the contribution of the state pair $(\bm{x},\bm{y})$ to the ensemble MPE\@. In particular, pairs with large negative values of $c_{t}^{\T b}(\bm{x},\bm{y})$ constitute major contributions to the ensemble MPE\@. Therefore, by calculating $c_{t}^{\T b}(\bm{x},\bm{y})$ from the average energy $\avet{\epsilon}{\T b}{\bm x}$ over trajectories of the cooling process, one can identify pairs of states that have dominant contributions to the ensemble MPE\@. We demonstrate this application below.

Our approach can be combined with an existing theoretical framework~\cite{Lu2017NonequilibriumThermodynamics} in two aspects, as detailed in SM~\cite{SM}.
First, Result 1 can be extended for another definition of ensemble MPE~\cite{Lu2017NonequilibriumThermodynamics} that uses the Kullback--Leibler (KL) divergence instead of the average energy. 
Second, focusing on the long-time behavior of Markovian systems, we can apply the eigenmode expansion in Ref.~\cite{Lu2017NonequilibriumThermodynamics} to $c_t^{\T b}(\bm x,\bm y)$. This enables us to evaluate the microscopic origin of ensemble MPEs based on the dominant eigenmodes instead of the average energy.

\paragraph{Result 2: No-Mpemba theorem}
Our second main result is a ``no-Mpemba theorem'' that rigorously proves the absence of ensemble MPEs for a broad class of systems. In Result 2, we restrict ourselves to discrete-state Markovian systems with a finite number of states. The transition rate from state $\bm{y}$ to $\bm{x}$ with a bath temperature $\T b$ is written as $R_{\T b}(\bm{x}\vert\bm{y})$, which determines the conditional distribution by
\begin{equation}
\frac{\partial }{\partial t} P_{t}^{\T b}(\bm{x}\vert\bm{y}) = \sum_{\bm{z}:\,\bm{z} \neq \bm{x}}\bigl[R_{\T b}(\bm{x}\vert\bm{z})P_{t}^{\T b}(\bm{z}\vert\bm{y})-R_{\T b}(\bm{z}\vert\bm{x})P_{t}^{\T b}(\bm{x}\vert\bm{y})\bigr].
\label{eq:master_equation}
\end{equation}
We assume that $R_{\T b}(\Bullet\vert\Bullet)$ is irreducible and has $\piT{\T b}(\Bullet)$ as the stationary distribution so that $\lim_{t \to \infty}P_{t}^{\T b}(\bm{x}\vert\bm{y}) = \piT{\T b}(\bm{x})$ from any $\bm{y}$~\cite{Norris1997markov}.
The main idea of Result 2, based on the theory of stochastic interacting systems~\cite{Liggett2005InteractingParticle,Levin2017MarkovChains}, is to introduce a partial ordering between states ($\bm{x} \preceq \bm{y}$), which reflects the system's inherent structures.

Result 2 is stated as follows: For a fixed range of temperature $[T_{1}, T_{2}]$, the ensemble MPE is absent for any $T_{1} \leq \T b < \T c < \T h \leq  T_{2}$ and any $t$ if there exists a partial ordering $\preceq$ with which the following conditions hold: \textit{(a) $\bm{x}\preceq\bm{y}$ implies $\epsilon(\bm{x}) \leq \epsilon(\bm{y})$};\textit{
(b) $R_{\T b}(\bm{x}\vert\bm{y})$ is nonzero only if $\bm{x}$ and $\bm{y}$ are {comparable} (i.e., $\bm{x}\preceq\bm{y}$ or $\bm{x}\succeq\bm{y}$) for any $\T b \in [T_{1}, T_{2}]$; (c) There exists a monotone coupling of $R_{\T b}(\Bullet\vert\Bullet)$ with respect to $\preceq$ \,for any $\T b \in [T_{1}, T_{2}]$.}

A monotone coupling of $R_{\T b}(\Bullet\vert\Bullet)$ is a fictitious Markovian process over two copies of the system $\mathcal{X}\times\mathcal{X}$ that satisfies the following two conditions~\cite{Liggett2005InteractingParticle,Levin2017MarkovChains}. 
First, letting $R_{\T b}^{\sharp}(\bm{u},\bm{v}\vert\bm{x},\bm{y})$ be the transition rate from $(\bm{x},\bm{y}) \in \mathcal{X}\times\mathcal{X}$ to $(\bm{u},\bm{v}) \in \mathcal{X}\times\mathcal{X}$ of the monotone coupling, $R_{\T b}^{\sharp}(\bm{u},\bm{v}\vert\bm{x},\bm{y})$ must be zero if $\bm{x}\preceq\bm{y}$ and $\bm{u}\not\preceq\bm{v}$. 
Second, the marginal process of each system copy must be identical to the original process $R_{\T b}(\Bullet\vert\Bullet)$. More precisely, we impose that $\sum_{\bm{v}}R_{\T b}^{\sharp}(\bm{u},\bm{v}\vert\bm{x},\bm{y}) = R_{\T b}(\bm{u}\vert\bm{x})$ holds independently of $\bm y$ for every $\bm{u},\bm{x}$ ($\bm{u} \neq \bm{x}$), and $\sum_{\bm{u}}R_{\T b}^{\sharp}(\bm{u},\bm{v}\vert\bm{x},\bm{y}) = R_{\T b}(\bm{v}\vert\bm{y})$ holds independently of $\bm {x}$ for every $\bm{v},\bm{y}$ ($\bm{v} \neq \bm{y}$).

Notably, the conditions (a)--(c) of the no-Mpemba theorem can be checked based on given $\epsilon(\Bullet)$, $R_{\T b}(\Bullet\vert\Bullet)$, $R_{\T b}^{\sharp}(\doubleBullet\vert\doubleBullet)$, and $\preceq$ without explicitly solving the Markovian time evolution. Therefore, this theorem applies to systems whose time evolution is not solvable. Finding suitable monotone coupling and partial ordering is not trivial, but we can use known constructions for many important stochastic models, including the stochastic Ising model, exclusion process, and contact process~\cite{Liggett2005InteractingParticle,Levin2017MarkovChains}. 

The proof of Result 2, found in SM~\cite{SM}, considers the monotone coupling with an initial state in $\mathcal{K} \coloneqq \{(\bm{x},\bm{y}) \mid \bm{x}\preceq\bm{y}\}$. 
Since $R_{\T b}^{\sharp}(\bm{u},\bm{v}\vert\bm{x},\bm{y})=0$ for $(\bm{x},\bm{y})\in \mathcal{K}$ and $(\bm{u},\bm{v})\notin \mathcal{K}$, the process stays in $\mathcal{K}$ indefinitely, meaning one copy does not overtake the other. 
This property is akin to the absence of MPEs and, when combined with classical theorems~\cite{Harris1977CorrelationInequality,Liggett2005InteractingParticle,Levin2017MarkovChains}, leads to Result 2.

\paragraph{Example 1: Ferromagnetic Ising models}
We demonstrate our two results with ferromagnetic Ising models, in which ensemble MPEs have been found by numerical calculations~\cite{Vadakkayil2021ShouldAHotterParamagnet,Ghosh2024SimulationsOfMpemba,Das2023PerspectivesOnAFewPuzzles,Chatterjee2024MpembaEffect}. 
Here, we consider the Ising model on an arbitrary lattice with $L$ sites under an arbitrary boundary condition.
A state of the Ising model is specified as $\bm{x} \equiv (x_{1},\dots,x_{L})$, where $x_{i} = \pm1$ denotes the direction of the spin at site $i$. We use the symbols $\mathord{\uparrow} \coloneqq +1$ and $\mathord{\downarrow} \coloneqq -1$. 
The number of states is $2^{L}$. 
The energy of the state $\bm{x}$ is $\epsilon(\bm{x}) = \sum_{i}H_{i}x_{i}-\frac{1}{2}\sum_{ij}J_{ij}x_{i}x_{j}$, where the magnetic fields $H_i$ and the coupling constants, $J_{ij} = J_{ji}\geq0$ and $J_{ii} = 0$, may depend on site indices. 
For any state~$\bm{x}$, $\bm{x}^{i} \coloneqq (x_{1},\dots,-x_{i},\dots,x_{L})$ denotes the state obtained by flipping the $i$th spin.
We assume that the transition rate from a state $\bm{x}$ to $\bm{x}^{i}$ is given by $R_{\T b}(\bm{x}^{i}\vert\bm{x}) = r_{i}^{\T b}(\epsilon(\bm{x}^{i})-\epsilon(\bm{x}))$, where $r_{i}^{\T b}(\Delta\epsilon)$ is an arbitrary nonzero decreasing function of $\Delta\epsilon$ that may depend on the site index $i$ and satisfies the detailed balance $r_{i}^{\T b}(\Delta\epsilon)/r_{i}^{\T b}(-\Delta\epsilon) = \exp(-\Delta\epsilon/\kB\T b)$. 
We set $R_{\T b}(\bm{x}\vert\bm{y}) = 0$ for any $\bm{x}$ and $\bm{y}$ whose spin configurations differ at two or more sites.
 
\begin{figure}
\includegraphics[width = 1\columnwidth]{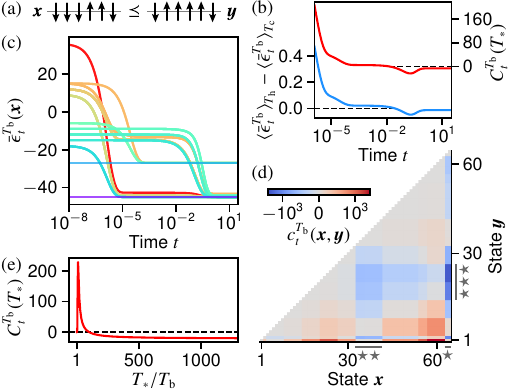}
\caption{Main results exemplified with ferromagnetic Ising models. 
(a) Partial ordering between states for applying the no-Mpemba theorem under strong magnetic fields. 
(b) Existence of the ensemble MPE when the magnetic field is weak, observed as the negativity of the energy difference $\aveTt{\epsilon}{\T b}{\T h}-\aveTt{\epsilon}{\T b}{\T c}$ for $(\kB\T h, \kB\T c) = (2000, 200)$ (blue) and, equivalently, the negativity of $C_{t}^{\T b}(T_{\largestar})$ for $\kB T_{\largestar} = 1000$ (red). We fix $\kB\T b = 1$ throughout. 
(c) Average energy $\avet{\epsilon}{\T b}{\bm{x}}$ starting with each initial microstate $\bm{x}$. Crossings between curves indicate the existence of microstate MPEs.
Colors indicate the initial state energy $\epsilon(\bm{x}) = \ave{\epsilon}{\T b}{\bm{x}}{0}$. 
(d) Measure of microstate MPE $c_{t}^{\T b}(\bm{x},\bm{y})$ at $t = 0.1$. The states $\bm{x}$ and $\bm{y}$ are sorted from the low-energy to the high-energy states and labeled with the integers from 1 to 64 for convenience; for example, 1 corresponds to the ground state $(\mathord{\downarrow}\mathord{\downarrow}\mathord{\downarrow}\mathord{\downarrow}\mathord{\downarrow}\mathord{\downarrow})$. 
Negative $c_{t}^{\T b}(\bm{x},\bm{y})$ occurs mainly in the three regions marked with gray \Star{} symbols. 
(e) By averaging $c_{t}^{\T b}(\bm{x},\bm{y})$ over $\piT{T_{\largestar}}(\bm{x})\piT{T_{\largestar}}(\bm{y})$, we obtain $C_{t}^{\T b}(T_{\largestar}) < 0$ for some $T_{\largestar}$, which is equivalent to the existence of the ensemble MPE\@. 
\label{fig:ising} }
\end{figure}

We use Result 2 to show that the ensemble MPE is absent when the magnetic field is strong, $H_{i} \geq \sum_{j} J_{ij}$ for all $i$. We use the partial ordering, $\bm{x}\preceq\bm{y}$ if and only if $x_{i} \leq  y_{i}$ for all $i = 1,\dots,L$ [Fig.~\ref{fig:ising}(a)],
and a monotone coupling explained in SM~\cite{SM}, both adopted from Ref.~\cite{Liggett2005InteractingParticle}. 
Then, the prerequisites (a)--(c) of Result 2 are satisfied for any $\T b$, and thus, the ensemble MPE is absent. We emphasize that this theorem applies to random parameters and random lattices.

On the other hand, ensemble MPEs may emerge when the magnetic field is weak, and their origins can be analyzed using Result 1. For concreteness, we consider a one-dimensional lattice with $L = 6$ under a periodic boundary condition. 
We set $H_{i} = 1.5$, $J_{i,i+1} = J_{i+1,i} = 6$ for all $i$ (we use $L+1\equiv1$) and any other coupling constants to zero. We use the transition rate $r_{i}^{\T b}(\Delta\epsilon) = \exp[-\Delta\epsilon/(2\kB\T b)]$ for all $i$. Figure~\ref{fig:ising}(b) shows that this system exhibits ensemble MPE. 

To explore its microscopic origin, we plot the average energy $\avet{\epsilon}{\T b}{\bm{x}}$ for each initial state $\bm{x}$ in Fig.~\ref{fig:ising}(c). The average energy shows a complicated behavior: Some states with high energy lose their energy quickly, while other states with intermediate levels of energy do not relax until later times. 
Thus, many pairs of states exhibit the microstate MPE\@.
In Fig.~\ref{fig:ising}(d), we plot $c_{t}^{\T b}(\bm{x},\bm{y})$ for all pairs at $t = 10^{-1}$, when the ensemble MPE is present. From the figure, we can identify three major regions of states that produce $c_{t}^{\T b}(\bm{x},\bm{y}) < 0$, marked with $\Star$ symbols. 
The region (\Star) consists of states with three isolated $\mathord{\uparrow}$ spins and three $\mathord{\downarrow}$ spins, and the region (\Star\Star) consists of states with four isolated $\mathord{\uparrow}$'s and two $\mathord{\downarrow}$'s. These states have high energy, but they can relax fast by flipping the isolated $\mathord{\uparrow}$ spins. 
The region (\Star\Star\Star) consists of the states $(\mathord{\uparrow}\mathord{\uparrow}\mathord{\uparrow}\mathord{\downarrow}\mathord{\downarrow}\mathord{\downarrow})$, $(\mathord{\uparrow}\mathord{\uparrow}\mathord{\uparrow}\mathord{\uparrow}\mathord{\downarrow}\mathord{\downarrow})$, and their cyclic permutations. 
Although these states have relatively low energy, they relax slowly because of their metastable domain $(\mathord{\uparrow}\mathord{\cdots}\mathord{\uparrow})$.
Thus, $\avet{\epsilon}{\T b}{\bm{x}}$ of these three regions crosses with $\avet{\epsilon}{\T b}{\bm{x}}$ of each other and of other states, leading to the microstate MPEs. 
Those negative $c_{t}^{\T b}(\bm{x},\bm{y})$ collectively produce $C_{t}^{\T b}(T_{\largestar}) < 0$, i.e., the ensemble MPE [Fig.~\ref{fig:ising}(e)]. Our Result 1 thus provides an intuitive microscopic explanation of ensemble MPEs that cannot be obtained otherwise. 

\paragraph{Example 2: One-dimensional particle system}
Another important class of many-body systems is particle systems. 
We consider an exclusion process on a one-dimensional lattice with reflective boundaries, a discretized model of particles with hard-core interactions. The lattice has $L$ sites, labeled as $1,\dots,L$ from left to right, and each site can accommodate up to one particle. 
The system contains $N$ ($ < L$) indistinguishable particles.
A state of the system is specified by the set of the occupied sites, $\bm{x} \equiv (x_{1},\dots,x_{N})$, in which $x_{i}$ is the position of the $i$th particle from the left satisfying $1 \leq x_{1} < \cdots < x_{N} \leq  L$\@. 
The energy of state $\bm{x}$ is $\epsilon(\bm{x}) = \sum_{i = 1}^{N}e_{x_{i}}$, where $e_{k}$ is the potential energy of site $k$. 
The particle at site $k$ jumps to the nearest neighbors $\ell = k\pm1$ with a nonzero rate $r_{\ell\vert k}^{\T b}$ if $1 \leq \ell \leq  L$ and $\ell$ is vacant. 
The rate does not depend on the configuration of the other particles, and it satisfies $r_{\ell \vert k}^{\T b}\exp(-e_{k}/\kB\T b) = r_{k\vert \ell}^{\T b}\exp(-e_{\ell}/\kB\T b)$. 

We use Result 2 to prove that this exclusion process does not show ensemble MPEs when the potential energy is monotone, $e_{1} \leq \cdots \leq e_{L}$. 
We follow Ref.~\cite{Levin2017MarkovChains} to introduce a partial ordering, $\bm{x}\preceq\bm{y}$ if and only if $x_{i} \leq  y_{i}$ for all $i = 1,...,N$, and a monotone coupling explained in SM~\cite{SM}. 
We can then prove the prerequisites (a)--(c) of Result 2 for any $\T b$ and, thus, the absence of the ensemble MPE. 

In the special case $N = 1$, this result states that a single particle system in a one-dimensional lattice with a monotone potential does not exhibit the ensemble MPE\@. By taking an appropriate continuum limit, such a system converges to a single particle system in a one-dimensional continuous space governed by a Langevin equation~\cite{kampen2007stochastic}. While we skip the mathematical rigor, we can thus see that a one-dimensional Langevin equation with a monotone potential does not exhibit ensemble MPEs, which is consistent with previous numerical results~\cite{Walker2021AnomalousThermal}. 
On the other hand, a single particle in a nonmonotonic potential can exhibit ensemble MPEs, and their microscopic origin can be analyzed using our Result 1 (see SM~\cite{SM}).

\paragraph{Discussion}
In this Letter, we have developed two universal microscopic theories of MPEs that rely on microstate comparisons. 
The microstate MPE provides a universal microscopic framework for understanding and analyzing ensemble MPEs, and the no-Mpemba theorem gives a general criterion for excluding the possibility of ensemble MPEs based on the microscopic description of the system.

Future work will apply these findings to a broader range of systems. Observing microstate MPEs in experiments and numerical simulations will help uncover the microscopic origins of MPEs and harness MPEs. Applying the no-Mpemba theorem to a broader class of systems will lead to a comprehensive classification of systems into those with and without MPEs. More generally, our results show the validity of rigorous and universal approaches to MPEs. It is also an interesting direction to connect our frameworks to another rigorous approach based on majorization~\cite{VuHayakawa}.

\vspace{\sectionsep} 
\begin{acknowledgments}
The authors thank Amit Kumar Chatterjee for stimulating discussion to initiate this project. N.O. and S.I. thank Tan Van Vu for sharing his preliminary manuscript with H. Hayakawa. N.O. thanks Hosho Katsura, Ryuna Nagayama, Ken Hiura, and Kohei Yoshimura for discussions. N.O. is supported by JSPS KAKENHI Grant No.\ 23KJ0732. H.H. is supported by the Kyoto University Foundation. S.I. is supported by JSPS KAKENHI Grants No.\ 21H01560, No.\ 22H01141, No.\ 23H00467, and No.\ 24H00834, JST ERATO Grant No.\ JPMJER2302, and UTEC-UTokyo FSI Research Grant Program.
\end{acknowledgments}

\onecolumngrid

\appendix 
\counterwithout{equation}{section} 

\clearpage 

\onecolumngrid 

\setcounter{equation}{0} 
\setcounter{figure}{0} 
\setcounter{table}{0} 
\setcounter{page}{1} 

\renewcommand{\theequation}{S\arabic{equation}} 
\renewcommand{\thefigure}{S\arabic{figure}} 
\renewcommand{\thesection}{\Alph{section}} 
\renewcommand{\thesubsection}{\arabic{subsection}} 

\let\oldsection\section 
\let\oldsubsection\subsection 
\let\olduppercase\uppercase 

\titleformat*{\section}{\centering\fontsize{10.5pt}{\baselineskip}\selectfont\bfseries} 
\titleformat*{\subsection}{\centering\fontsize{10.5pt}{\baselineskip}\selectfont\bfseries} 
\titleformat*{\subsubsection}{\centering\fontsize{10.5pt}{\baselineskip}\selectfont\itshape}  

\titlespacing*{\section}{\linewidth}{3em}{0.4em} 
\titlespacing*{\subsection}{\linewidth}{1.5em}{0.4em} 
\titlespacing*{\subsubsection}{\linewidth}{1.5em}{0.4em} 

\fontsize{10.5pt}{12.0pt}\selectfont 
\fontdimen2\font = 2.625pt

\begin{center} 
\textbf{\large Supplemental Material for\\[0.1em]
Microscopic theory of Mpemba effects and\\[0.1em]
a no-Mpemba theorem for monotone many-body systems}

\vspace*{1em}
Naruo\hspace{0.35em}Ohga,\hspace{0.35em}Hisao\hspace{0.35em}Hayakawa,\hspace{0.35em}and\hspace{0.35em}Sosuke\hspace{0.35em}Ito 

\vspace*{1.5em} 
\end{center}

This Supplemental Material is organized as follows. 
We prove Result 1 in Sec.~\ref{sec:result1}\@.
In Sec~\ref{sec:entropic}, we extend Result 1 to another definition of ensemble MPEs based on the KL divergence. In Sec.~\ref{sec:eigenmode}, we relate Result 1 with the eigenmode analysis of Markovian systems.
We prove Result 2 in Sec.~\ref{sec:result2} and provide the details of the application of Result 2 to ferromagnetic Ising models and exclusion processes in Sec.~\ref{sec:result2_application}\@. 
Section~\ref{sec:numerics} provides supplemental numerical results, including a numerical test of the no-Mpemba theorem. 

\vspace*{1.5em}
\twocolumngrid

\section{Proof of Result 1 
\label{sec:result1}}

We prove Result 1. We first prepare the following Lemma.

\paragraph{Lemma S1}
The quantity $C_{t}^{\T b}(T_{\largestar})$ is related to the derivative of $\aveTt{\epsilon}{\T b}{\T i}$ with respect to the initial temperature $\T i$: 
\begin{equation}
C_{t}^{\T b}(T_{\largestar}) = 2\kB T_{\largestar}^{2}\left.\left(\frac{\partial\aveTt{\epsilon}{\T b}{\T i}}{\partial\T i}\right)\right|_{\T i = T_{\largestar}}.
\label{pr:Mpemba_derivative}
\end{equation}

\paragraph{Proof of Lemma S1}
We first calculate the derivative of the initial Gibbs distribution:
\begin{align}
\frac{\partial\piT{\T i}(\bm{y})}{\partial\T i} &  = \piT{\T i}(\bm{y})\frac{\partial}{\partial\T i}\ln\piT{\T i}(\bm{y})\nonumber \\
 &  = \piT{\T i}(\bm{y})\left[\frac{\epsilon(\bm{y})}{\kB\T i^{2}}-\frac{\intX x\,e^{-\epsilon(\bm{x})/\kB\T i}\epsilon(\bm{x})/\kB\T i^{2}}{\intX x\,e^{-\epsilon(\bm{x})/\kB\T i}}\right]\nonumber \\
 &  = \frac{1}{\kB\T i^{2}}\piT{\T i}(\bm{y})\left[\epsilon(\bm{y})-\langle\epsilon\rangle_{\T i}\right],
\label{pr:temperature}
\end{align}
where we use $\langle\epsilon\rangle_{\T i} \coloneqq \intX x\,\epsilon(\bm{x})\piT{\T i}(\bm{x})$. Using this expression, the $\T i$-derivative of $\aveTt{\epsilon}{\T b}{\T i}$ is given by
\begin{align}
\frac{\partial\aveTt{\epsilon}{\T b}{\T i}}{\partial\T i} &  = \frac{\partial}{\partial\T i}\intX y\,\avet{\epsilon}{\T b}{\bm{y}}\piT{\T i}(\bm{y})\nonumber \\
 &  = \frac{1}{\kB\T i^{2}}\intX y\,\avet{\epsilon}{\T b}{\bm{y}}\left[\epsilon(\bm{y})-\langle\epsilon\rangle_{\T i}\right]\piT{\T i}(\bm{y}).
\label{pr:FDR}
\end{align}
The last expression of Eq.~\eqref{pr:FDR} is the correlation function between the energy at time $0$ and the average energy at time $t$, and thus Eq.~\eqref{pr:FDR} is essentially the fluctuation--dissipation theorem. 
We rearrange Eq.~\eqref{pr:FDR} to obtain
\begin{align}
 & 2\kB\T i^{2}\frac{\partial\aveTt{\epsilon}{\T b}{\T i}}{\partial\T i}\nonumber \\
 &  = 2\intXX xy\,\avet{\epsilon}{\T b}{\bm{y}}[\epsilon(\bm{y})-\epsilon(\bm{x})]\piT{\T i}(\bm{x})\piT{\T i}(\bm{y})\nonumber \\
 &  = \intXX xy\,\bigl[\avet{\epsilon}{\T b}{\bm{y}}-\avet{\epsilon}{\T b}{\bm{x}}\bigr][\epsilon(\bm{y})-\epsilon(\bm{x})]\piT{\T i}(\bm{x})\piT{\T i}(\bm{y})\nonumber \\
 &  = C_{t}^{\T b}(\T i),
\label{pr:rearrange}
\end{align}
where the first equality uses $\intX x\,\piT{\T i}(\bm{x}) = 1$, and the second equality follows from the exchange of variables $\bm{x}$ and $\bm{y}$. This proves Eq.~\eqref{pr:Mpemba_derivative}. \qed

Using this Lemma, we prove Result 1.

\paragraph{Proof of Result 1}
The average energy $\aveTt{\epsilon}{\T b}{\T i}$ is continuously differentiable in $\T i$. This follows from the fact that the right-hand side of Eq.~\eqref{pr:FDR} is continuous in $\T i$. 

Suppose that the ensemble MPE $\aveTt{\epsilon}{\T b}{\T c} > \aveTt{\epsilon}{\T b}{\T h}$ is present for a pair of temperatures $\T c < \T h$. Then, the mean-value theorem implies that there exists a temperature $T_{\largestar}$ such that $\T c < T_{\largestar} < \T h$ and $\bigl(\partial\aveTt{\epsilon}{\T b}{\T i}/\partial\T i\bigr)\bigr\vert_{\T i = T_{\largestar}} < 0$. Combined with the Lemma S1, there exists a temperature $T_{\largestar}$ such that $\T c < T_{\largestar} < \T h$ and $C_{t}^{\T b}(T_{\largestar}) < 0$.

Conversely, suppose that $C_{t}^{\T b}(T_{\largestar}) < 0$ holds for some $T_{\largestar} > \T b$. Then, Lemma S1 implies that $\bigl(\partial\aveTt{\epsilon}{\T b}{\T i}/\partial\T i\bigr)\bigr\vert_{\T i = T_{\largestar}} < 0$. Since $\partial\aveTt{\epsilon}{\T b}{\T i}/\partial\T i$ is continuous in $\T i$, we can take $\T c$ and $\T h$ such that $\T b < \T c < T_{\largestar} < \T h$ and $\partial\aveTt{\epsilon}{\T b}{\T i}/\partial\T i < 0$ for all $\T i \in [\T c,\T h]$. Over the interval $\T i \in [\T c,\T h]$, $\aveTt{\epsilon}{\T b}{\T i}$ is monotonically decreasing in $\T i$. Therefore, $\aveTt{\epsilon}{\T b}{\T c} > \aveTt{\epsilon}{\T b}{\T h}$ holds, i.e., the ensemble MPE is present. \qed

\section{Variant of Result 1 with the KL divergence 
\label{sec:entropic}}

\subsection{The KL divergence and nonequilibrium free energy}

Some literature defines the ensemble MPE using quantities other than the average energy. One of the often-used quantities is the KL divergence between the instantaneous probability distribution and the final Gibbs distribution, also called the \textit{entropic distance}
in Ref.~\cite{Lu2017NonequilibriumThermodynamics}. Here, we define it in an equivalent but seemingly different way. We first introduce the instantaneous probability distribution of the relaxation process starting with a Gibbs distribution $\piT{\T i}(\Bullet)$, $p_{t}^{\T b}(\bm{x};\T i) \coloneqq \intX y\,P_{t}^{\T b}(\bm{x}\vert\bm{y})\piT{\T i}(\bm{y})$. We then define the stochastic entropy $s_{t}^{\T b}(\bm{x};\T i) \coloneqq -\kB\ln p_{t}^{\T b}(\bm{x};\T i)$ and the nonequilibrium free energy $f_{t}^{\T b}(\bm{x};\T i) \coloneqq \epsilon(\bm{x})-\T bs_{t}^{\T b}(\bm{x};\T i)$. 

We consider the nonequilibrium free energy as an explicitly time-dependent observable. More precisely, we regard $f_{t}^{\T b}(\bm{x};\T i)$ as the value of the observable when the system is in state $\bm{x}$ at time $t$. This formulation is standard in stochastic thermodynamics (see, e.g., Eqs.~({60})--({63}) of Ref.~\cite{VandenBroeck2015EnsembleAndTrajectory}). The average of the nonequilibrium free energy at time $t$ over the relaxation process starting with a single fixed initial state $\bm{y} \in \mathcal{X}$ is given by $\bar{f}_{t}^{\T b}(\bm{y};\T i) \coloneqq \intX x\,f_{t}^{\T b}(\bm{x};\T i)P_{t}^{\T b}(\bm{x}\vert\bm{y})$. Note that $\bar{f}_{t}^{\T b}(\bm{x};\T i)$ is not equal to $f_{t}^{\T b}(\bm{x};\T i)$ in general while $\bar{f}_{0}^{\T b}(\bm{x};\T i) = f_{0}^{\T b}(\bm{x};\T i)$ holds at time 0. Similarly, the average of the nonequilibrium free energy at time $t$ over the relaxation process starting from the Gibbs distribution $\piT{\T i}(\Bullet)$ is $\langle\bar{f}_{t}^{\T b}\rangle_{\T i} \coloneqq \intX y\,\bar{f}_{t}^{\T b}(\bm{y};\T i)\piT{\T i}(\bm{y})$, and it can be rewritten as $\langle\bar{f}_{t}^{\T b}\rangle_{\T i} = \intX x\,f_{t}^{\T b}(\bm{x};\T i)p_{t}^{\T b}(\bm{x};\T i)$. 

Using these definitions, we define the entropic distance as $D_{t}^{\T b}(\T i) \coloneqq \langle\bar{f}_{t}^{\T b}\rangle_{\T i}-F_{\T b}$, where $F_{\T b}$ is the equilibrium free energy introduced in the main text. This turns out to be equal to $\kB\T b$ times the expression of the entropic distance in Ref.~\cite{Lu2017NonequilibriumThermodynamics}:
\begin{align}
D_{t}^{\T b}(\T i) &  = \intX x\,\epsilon(\bm{x})[p_{t}^{\T b}(\bm{x};\T i)-\piT{\T b}(\bm{x})]\nonumber \\
 & \quad+\kB\T b\intX x\,p_{t}^{\T b}(\bm{x};\T i)\ln p_{t}^{\T b}(\bm{x};\T i)\nonumber \\
 & \quad-\kB\T b\intX x\,\piT{\T b}(\bm{x})\ln\piT{\T b}(\bm{x}).
\label{ent:equal_entropic}
\end{align}
It is also equal to $\kB\T b$ times the KL divergence between the instantaneous distribution and the Gibbs distribution:
\begin{equation}
D_{t}^{\T b}(\T i) = \kB\T b\intX x\,p_{t}^{\T b}(\bm{x};\T i)\ln\frac{p_{t}^{\T b}(\bm{x};\T i)}{\piT{\T b}(\bm{x})}.
\label{ent:equal_KL}
\end{equation}

\paragraph{Derivation of Eqs.~\eqref{ent:equal_entropic} and \eqref{ent:equal_KL}}
We insert the definition of $f_{t}^{\T b}(\bm{x};\T i)$ into $\langle\bar{f}_{t}^{\T b}\rangle_{\T i} = \intX x\,f_{t}^{\T b}(\bm{x};\T i)p_{t}^{\T b}(\bm{x};\T i)$ to obtain 
\begin{align}
D_{t}^{\T b}(\T i) &  = \intX x\,\epsilon(\bm{x})p_{t}^{\T b}(\bm{x};\T i)\nonumber \\
 & \quad+\kB\T b\intX x\,p_{t}^{\T b}(\bm{x};\T i)\ln p_{t}^{\T b}(\bm{x};\T i)-F_{\T b}.
\label{ent:rewrite}
\end{align}
Averaging $\kB\T b \ln\piT{\T b}(\bm{x}) = - \epsilon(\bm{x}) + F_{\T b}$ over $\piT{\T b}(\bm{x})$ gives
\begin{equation}
F_{\T b} = \kB\T b\intX x\,\piT{\T b}(\bm{x})\ln\piT{\T b}(\bm{x})+\intX x\,\epsilon(\bm{x})\piT{\T b}(\bm{x}).
\end{equation}
Inserting this relation to $F_{\T b}$ in Eq.~\eqref{ent:rewrite} gives Eq.~\eqref{ent:equal_entropic}. On the other hand, averaging $\kB\T b \ln\piT{\T b}(\bm{x}) = - \epsilon(\bm{x}) + F_{\T b}$ over $p_t^{\T b}(\bm{x};\T i)$ leads to
\begin{align}
 & \intX x\,\epsilon(\bm{x})p_{t}^{\T b}(\bm{x};\T i)-F_{\T b}\nonumber \\
 & \quad = -\kB\T b\intX x\,p_{t}^{\T b}(\bm{x};\T i)\ln\piT{\T b}(\bm{x}).
\end{align}
Inserting this expression into Eq.~\eqref{ent:rewrite} gives Eq.~\eqref{ent:equal_KL}.
\qed

\subsection{Variant of Result 1}

Since it is shown that $D_{0}^{\T b}(\T h) > D_{0}^{\T b}(\T c)$ at time 0, the condition 
\begin{equation}
D_{t}^{\T b}(\T h) < D_{t}^{\T b}(\T c)
\label{ent:ensemble_def}
\end{equation}
has been used to define the ensemble MPE~\cite{Lu2017NonequilibriumThermodynamics}. We define the corresponding microstate MPE by the negativity of 
\begin{align}
 & \tilde{c}_{t}^{\T b}(\bm{x},\bm{y};T_{\largestar})\nonumber \\
 &  \coloneqq \bigl[\bar{f}_{t}^{\T b}(\bm{x};\T{\largestar})-\bar{f}_{t}^{\T b}(\bm{y};\T{\largestar})\bigr]\bigl[f_{0}^{\T b}(\bm{x};\T{\largestar})-f_{0}^{\T b}(\bm{y};\T{\largestar})\bigr]
 \label{ent:def_microstate}
\end{align}
for a pair of states $\bm{x}$ and $\bm{y}$. Since $\bar{f}_{0}^{\T b}(\bm{x};\T{\largestar}) = f_{0}^{\T b}(\bm{x};\T{\largestar})$ and $\bar{f}_{0}^{\T b}(\bm{y};\T{\largestar}) = f_{0}^{\T b}(\bm{y};\T{\largestar})$ at time $0$, this definition captures the crossing between the curves of $\bar{f}_{t}^{\T b}(\bm{x};\T{\largestar})$ and $\bar{f}_{t}^{\T b}(\bm{y};\T{\largestar})$, similarly to Fig.~\ref{fig:concept}(b) in the main text. With these alternative definitions, the following variant of Result 1 holds:

\paragraph{Result S2}
For fixed $\T b$ and $t$, the following two statements are equivalent: \textit{(a) The entropic distances satisfy $D_{t}^{\T b}(\T h) < D_{t}^{\T b}(\T c)$ at time $t$ for at least one pair of initial temperatures $\T h > \T c$ ($ > \T b$); (b) The average of $\tilde{c}_{t}^{\T b}(\bm{x},\bm{y};\T{\largestar})$,
\begin{equation}
\tilde{C}_{t}^{\T b}(T_{\largestar}) \coloneqq \intXX xy\,\tilde{c}_{t}^{\T b}(\bm{x},\bm{y};T_{\largestar})\piT{T_{\largestar}}(\bm{x})\piT{T_{\largestar}}(\bm{y})
\label{ent:main_entropic}
\end{equation}
is negative for at least one temperature $T_{\largestar}$ ($ > \T b$).}
Moreover, the temperatures appearing in these statements can be taken to satisfy $\T{c} < T_{\largestar} < \T h$.

\vspace{\sectionsep}

\begin{figure}
\includegraphics[width = 1\columnwidth]{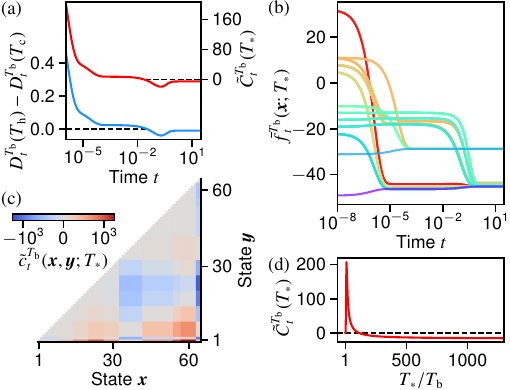}
\caption{Variant of Result 1 with the KL divergence (entropic distance), exemplified with the ferromagnetic Ising model in the main text. All parameter values are the same as in Fig.~\ref{fig:ising}. (a) Existence of ensemble MPE, shown by a negative difference in entropic distances $D_{t}^{\T b}(\T h)-D_{t}^{\T b}(\T c)$ and a negative value of $\tilde{C}_{t}^{\T b}(T_{\largestar})$. (b) Average free energy $\bar{f}_{t}^{\T b}(\bm{x};T_{\largestar})$ starting with each microstate $\bm{x}$. Crossings of curves indicate the existence of microstate MPEs. (c) Measure of microstate MPE $\tilde{c}_{t}^{\T b}(\bm{x},\bm{y};T_{\largestar})$ at $t = 0.1$. The positions of negative $\tilde{c}_{t}^{\T b}(\bm{x},\bm{y}; T_{\largestar})$ are similar to those in Fig.~\ref{fig:ising}(d). (d) The average of the measure of microstate MPE takes negative values, implying the existence of the ensemble MPE. 
\label{fig:ising_freeenergy}}
\end{figure}

Similarly to the discussion for Result 1 in the main text, this result implies that we can interpret a pair of state $(\bm x,\bm y)$ with a large negative value of $\tilde{c}_t^{\T b}(\bm x, \bm y; T_\largestar)$ as a major origin of the ensemble MPE defined with the KL divergence. We exemplify this finding in Fig.~\ref{fig:ising_freeenergy} using the same ferromagnetic Ising model as in the main text. All plots in Fig.~\ref{fig:ising_freeenergy} are qualitatively similar to those in Fig.~\ref{fig:ising}. Thus, Result S2 can capture the microscopic origin of the ensemble MPE similarly to Result 1, showing the robustness of our approach.

The proof of Result S2 is similar to the proof of Result 1. We start with the following Lemma.

\paragraph{Lemma S3}
The quantity $\tilde{C}_{t}^{\T b}(T_{\largestar})$ is related to the derivative of $D_{t}^{\T b}(\T i)$ with respect to the initial temperature $\T i$:
\begin{equation}
\tilde{C}_{t}^{\T b}(T_{\largestar}) = 2\kB T_{\largestar}^{2}\left(1-\frac{\T b}{T_{\largestar}}\right)\left.\left(\frac{\partial}{\partial\T i}D_{t}^{\T b}(\T i)\right)\right|_{\T i = T_{\largestar}}.
\label{ent:D_is_B}
\end{equation}
Note that the coefficient $2\kB T_\largestar^2 (1-\T b/T_\largestar)$ is positive for $T_\largestar > \T b$.

\paragraph{Proof of Lemma S3}
The derivative of $D_{t}^{\T b}(\T i)$ reads
\begin{align}
\frac{\partial}{\partial\T i}D_{t}^{\T b}(\T i) &  = \frac{\partial}{\partial\T i}\intX x\,f_{t}^{\T b}(\bm{x};\T i)p_{t}^{\T b}(\bm{x};\T i)\nonumber \\
 &  = \intX x\,f_{t}^{\T b}(\bm{x};\T i)\frac{\partial p_{t}^{\T b}(\bm{x};\T i)}{\partial\T i},
\label{ent:derivation1}
\end{align}
where the second equality follows from 
\begin{align}
 & \intX x\,\frac{\partial f_{t}^{\T b}(\bm{x};\T i)}{\partial\T i}p_{t}^{\T b}(\bm{x};\T i) \nonumber \\
 & \qquad = \kB\T b\intX x\,p_{t}^{\T b}(\bm{x};\T i)\frac{\partial}{\partial\T i}\ln p_{t}^{\T b}(\bm{x};\T i)\nonumber \\
 & \qquad = \kB\T b\frac{\partial}{\partial\T i}\intX x\,p_{t}^{\T b}(\bm{x};\T i)\nonumber \\
 & \qquad = 0.
\end{align}
We then calculate the derivative of $p_{t}^{\T b}(\bm{x};\T i)$:
\begin{align}
 & \frac{\partial p_{t}^{\T b}(\bm{x};\T i)}{\partial\T i}\nonumber \\
 &  = \intX y\,P_{t}^{\T b}(\bm{x}\vert\bm{y})\frac{\partial\piT{\T i}(\bm{y})}{\partial\T i}\nonumber \\
 &  = \frac{1}{\kB \T{i}^{2}}\intX y\,P_{t}^{\T b}(\bm{x}\vert\bm{y})\piT{\T i}(\bm{y})\left[\epsilon(\bm{y})-\langle\epsilon\rangle_{\T i}\right]\nonumber \\
 &  = \frac{\left(1-\T b/\T i\right)^{-1}}{\kB \T{i}^{2}}\nonumber \\
 & \quad\times\intX y\,P_{t}^{\T b}(\bm{x}\vert\bm{y})\piT{\T i}(\bm{y})\left[f_{0}^{\T b}(\bm{y};\T i)-\langle f_{0}^{\T b}\rangle_{\T i}\right],
\label{ent:derivation2}
\end{align}
where we define $\langle f_{0}^{\T b}\rangle_{\T i} \coloneqq \intX y\,f_{0}^{\T b}(\bm{y};\T i)\piT{\T i}(\bm{y})$. Here, the second equality uses Eq.~\eqref{pr:temperature}, and the last equality follows from 
\begin{align}
f_{0}^{\T b}(\bm{y};\T i) &  = \epsilon(\bm{y})+\kB\T b\ln\piT{\T i}(\bm{y})\nonumber \\
 &  = \epsilon(\bm{y})-\frac{\kB\T b}{\kB\T i}[\epsilon(\bm{y})-F_{\T i}]\nonumber \\
 &  = \left(1-\frac{\T b}{\T i}\right)\epsilon(\bm{y})+\frac{\T b}{\T i}F_{\T i}
 \label{ent:f_is_e1}
\end{align}
and 
\begin{align}
\langle f_{0}^{\T b}\rangle_{\T i} &  = \left(1-\frac{\T b}{\T i}\right)\intX y\,\epsilon(\bm{y})\piT{\T i}(\bm{y})+\frac{\T b}{\T i}F_{\T i}\nonumber \\
 &  = \left(1-\frac{\T b}{\T i}\right)\langle\epsilon\rangle_{\T i}+\frac{\T b}{\T i}F_{\T i}.
 \label{ent:f_is_e2}
\end{align}
Inserting Eq.~\eqref{ent:derivation2} into Eq.~\eqref{ent:derivation1} gives
\begin{align}
 \frac{\partial}{\partial\T i}D_{t}^{\T b}(\T i) = \frac{\left(1-\T b/\T i\right)^{-1}}{\kB \T{i}^{2}} \intXX xy\, \biggl\{ f_t^{\T b}(\bm{x};\T i)\,  \nonumber \\
 \phantom{A} \times P_{t}^{\T b}(\bm{x}\vert\bm{y}) \piT{\T i}(\bm{y})\left[f_{0}^{\T b}(\bm{y};\T i)-\langle f_{0}^{\T b}\rangle_{\T i}\right] \biggr\}.
\end{align}
Using the definition of ${\bar f}_t^{\T b}(\bm{y};\T i)$, we obtain
\begin{align}
 & \frac{\partial}{\partial\T i}D_{t}^{\T b}(\T i) = \frac{\left(1-\T b/\T i\right)^{-1}}{\kB \T{i}^{2}}\nonumber \\
 & \quad \times \intX y\,\bar{f}_{t}^{\T b}(\bm{y};\T i)\left[f_{0}^{\T b}(\bm{y};\T i)-\langle f_{0}^{\T b}\rangle_{\T i}\right]\piT{\T i}(\bm{y}).
\label{ent:derivation_unite}
\end{align}
This expression has the same structure as Eq.~\eqref{pr:FDR} with the replacement of $\avet{\epsilon}{\T b}{\bm{y}} \to \bar{f}_{t}^{\T b}(\bm{y};\T i)$, $\epsilon(\bm{y}) \to  f_{0}^{\T b}(\bm{y};\T i)$, and $\kB\T{i}^{2} \to (1-\T b/\T i)\kB\T i^{2}$. Thus, rearranging this expression similarly to Eq.~\eqref{pr:rearrange} gives 
\begin{equation}
2\left(1-\frac{\T b}{\T i}\right)\kB\T{i}^{2}\frac{\partial}{\partial\T i}D_{t}^{\T b}(\T i) = \tilde{C}_{t}^{\T b}(\T i).
\end{equation}
This proves Eq.~\eqref{ent:D_is_B}. \qed

\paragraph{Proof of Result S2}
We first verify that $D_{t}^{\T b}(\T i)$ is continuously differentiable with respect to $\T i$ for $\T i > \T b$. The nonequilibrium free energy $f_{t}^{\T b}(\bm{x};\T i)$ is a continuous function of $\T i$, as follows from its definition. Therefore, the averages $\bar{f}_{t}^{\T b}(\bm{y};\T i)$ and $\langle f_{0}^{\T b}\rangle_{\T i}$ are also continuous in $\T i$. Thus, the right-hand side of Eq.~\eqref{ent:derivation_unite} is continuous in $\T i$ for $\T i > \T b$, which means that $D_{t}^{\T b}(\T i)$ is continuously differentiable.

Using this fact, Result S2 is proven similarly to the proof of Result 1 in Appendix~\ref{sec:result1}\@. We omit further details. \qed

\section{Relation to Markovian eigenmode analysis 
\label{sec:eigenmode}}

\subsection{Eigenmodes of Markovian time evolution}

In this section, we focus on Markovian systems and relate our Result 1 to the eigenmode analysis of the Markovian generator of time evolution. Let $\mathcal{L}_{\T b}$ denote the generator of the Markovian time evolution of the probability distribution with the bath temperature $\T b$. The generator is mathematically a linear map from a function on $\mathcal{X}$ to another function on $\mathcal{X}$. For example, $\mathcal{L}_{\T b}$ may be a Fokker--Planck operator for continuous systems, for which $\mathcal{L}_{\T b}P_{t}^{\T b}(\Bullet\vert\bm{y})$ gives the right-hand side of the Fokker--Planck equation, such as the one in Eq.~\eqref{num:Fokker-Planck}. For discrete-state systems, $\mathcal{L}_{\T b}$ is the generator of a Markov jump process, and $\mathcal{L}_{\T b}P_{t}^{\T b}(\Bullet\vert\bm{y})$ gives the right-hand side of the master equation in Eq.~\eqref{eq:master_equation}. We define the adjoint operator $\mathcal{L}_{\T b}^{\dagger}$ through the identity $\intX x\,a(\bm{x})\mathcal{L}_{\T b}b(\bm{x}) = \intX x\,b(\bm{x})\mathcal{L}_{\T b}^{\dagger}a(\bm{x})$ for arbitrary real-valued functions $a(\bm{x})$ and $b(\bm{x})$ in the function space on~$\mathcal{X}$.

We review the basic properties of the eigenvalues and eigenfunctions of $\mathcal{L}_{\T b}$ and $\mathcal{L}_{\T b}^{\dagger}$ (e.g., Chapter 6 of Ref.~\cite{Risken1989FokkerPlanck}). The eigenvalue equations for $\mathcal{L}_{\T b}$ and $\mathcal{L}_{\T b}^{\dagger}$ read
\begin{equation}
\mathcal{L}_{\T b}\phi_{\alpha}^{\T b}(\bm{x}) = \lambda_{\alpha}^{\T b}\phi_{\alpha}^{\T b}(\bm{x}),\quad\mathcal{L}_{\T b}^{\dagger}\psi_{\alpha}^{\T b}(\bm{x}) = \lambda_{\alpha}^{\T b}\psi_{\alpha}^{\T b}(\bm{x}),
\end{equation}
where $\phi_{\alpha}^{\T b}(\bm{x})$ and $\psi_{\alpha}^{\T b}(\bm{x})$ are the eigenfunctions. We assume that the eigenvalues are discrete, labeled by $\alpha = 1,2\dots$. The eigenfunctions can be taken to be biorthonormal $\intX x\,\phi_{\alpha}^{\T b}(\bm{x})\psi_{\beta}^{\T b}(\bm{x}) = \delta_{\alpha\beta}$. For simplicity, we further assume that the sets of eigenfunctions $\{\phi_{\alpha}^{\T b}\}$ and $\{\psi_{\alpha}^{\T b}\}$ each form a complete set of basis of the function space on $\mathcal{X}$. Then, the conditional probability is expanded as 
\begin{equation}
P_{t}^{\T b}(\bm{x}\vert\bm{y}) = \sum_{\alpha\geq1}\exp(\lambda_{\alpha}^{\T b}t)\phi_{\alpha}^{\T b}(\bm{x})\psi_{\alpha}^{\T b}(\bm{y}).
\label{eig:expansion}
\end{equation}
We also assume for simplicity that all the eigenvalues and eigenfunctions are real, which is the case for overdamped systems in contact with a single reservoir, and sort the eigenvalues as $\lambda_{1}^{\T b} \geq \lambda_{2}^{\T b} \geq \cdots$. It is known that the first eigenvalue and eigenfunctions satisfy $\lambda_{1}^{\T b} = 0$, $\phi_{1}^{\T b}(\bm{x}) = \piT{\T b}(\bm{x})$, and $\psi_{1}^{\T b}(\bm{x}) = 1$. Thus, all eigenvalues are nonpositive, and $\intX x \, \piT{\T b} (\bm x) \psi_\alpha^{\T b}(\bm x) = \intX x\,\phi_\alpha^{\T b}(\bm x) = 0$ holds for all $\alpha \geq 2$, as follows from the biorthonormality.

\subsection{Eigenmode analysis of ensemble MPEs}

Ensemble MPEs in Markovian systems have been commonly analyzed in terms of eigenmodes~\cite{Lu2017NonequilibriumThermodynamics,Israel2019PRX,Busiello2021InducingAndOptimizing,Gal2020Precooling,Cao2022FastFunctionalization,Chtrite2021TheMetastableMpemba,Walker2021AnomalousThermal,Walker2022MpembaEffect,Biswas2023MpembaEffect1,Biswas2023MpembaEffect2,Teza2023RelaxationShortcuts,Teza2023EigenvalueCrossing,Pemartin2024ShortcutsOf,Chatterjee2023QuantumMpemba,Chatterjee2024Multiple,Shapira2024Inverse,Ivander2023Hyperacceleration,Carollo2021ExponentiallyAccelerated,Kochsiek2022Accelerating,Wang2024MpembaEffects,Strachan2024NonMarkovianQuantum,Kumar2020ExponentiallyFaster,Kumar2022AnomalousHeating,Nava2024MpembaEffects,Chittari2023GeometricApproach,Moroder2024ThermodynamicsOfTheQuantum,Deguenther2022AnomalousRelaxation}. 
While the analysis is valid for any Markovian system, it becomes the most relevant when a few modes are dominant for the time evolution in the focused time region. As has been done in literature, we assume $0 = \lambda_{1}^{\T b} > \lambda_{2}^{\T b} > \lambda_{3}^{\T b}$ and consider a large enough $t$ so that $\exp(\lambda_{3}^{\T b}t),\exp(\lambda_{4}^{\T b}t),\ldots$ are exponentially smaller than $\exp(\lambda_{2}^{\T b}t)$. We can approximate Eq.~\eqref{eig:expansion} as 
\begin{equation}
P_{t}^{\T b}(\bm{x}\vert\bm{y}) \simeq \piT{\T b}(\bm{x})+\exp(\lambda_{2}^{\T b}t)\phi_{2}^{\T b}(\bm{x})\psi_{2}^{\T b}(\bm{y})    
\label{eig:expansion_asymptotic}
\end{equation}
in this asymptotic regime.

Reference~\cite{Lu2017NonequilibriumThermodynamics} defined the ensemble MPEs in terms of the KL divergence [Eq.~\eqref{ent:ensemble_def}] and showed that the ensemble MPE is present in this asymptotic regime if 
\begin{equation}
\bigl\vert a_{2}^{\T b}(\T h) \bigr\vert < \bigl\vert a_{2}^{\T b}(\T c) \bigr\vert.
\label{eig:condition_RuLaz}
\end{equation}
Here, $a_{2}^{\T b}(\T i) \coloneqq \intX y\,\psi_{2}^{\T b}(\bm{y})\piT{\T i}(\bm{y})$ is the expansion coefficient corresponding to the second eigenfunction when we expand $\piT{\T i}(\Bullet)$ in terms of $\{\phi_{\alpha}^{\T b}(\Bullet)\}$. Moreover, the condition in Eq.~\eqref{eig:condition_RuLaz} is valid for a broader range of definitions of ensemble MPEs~\cite{Lu2017NonequilibriumThermodynamics}. However, it does not immediately apply to our definition of ensemble MPEs in Eq.~\eqref{eq:ensemble_def}.

Our definition of ensemble MPEs in terms of the average energy is related to the eigenfunctions in a different manner. Using Eq.~\eqref{eig:expansion_asymptotic}, we can expand the average energy in the asymptotic regime as
\begin{equation}
\aveTt{\epsilon}{\T b}{\T i}
  \simeq \langle\epsilon\rangle_{\T b}+\exp(\lambda_{2}^{\T b}t)\epsilon_{2}^{\T b}a_{2}^{\T b}(\T i),
\label{eig:energy_evolution}
\end{equation}
where we define $\epsilon_{2}^{\T b} \coloneqq \intX x\,\epsilon(\bm{x})\phi_{2}^{\T b}(\bm{x})$, which is the expansion coefficient on the second eigenfunction when we expand $\epsilon(\Bullet)$ in terms of $\{\psi_{\alpha}^{\T b}(\Bullet)\}$. The difference of the average energies from two initial temperatures $\T i = \T h,\T c$ reads
\begin{equation}
\aveTt{\epsilon}{\T b}{\T h}-\aveTt{\epsilon}{\T b}{\T c} \simeq \exp(\lambda_{2}^{\T b}t)\epsilon_{2}^{\T b}\left[a_{2}^{\T b}(\T h)-a_{2}^{\T b}(\T c)\right].
\end{equation}
Thus, the ensemble MPE is present in the asymptotic regime if the condition
\begin{equation}
\epsilon_{2}^{\T b}a_{2}^{\T b}(\T h) < \epsilon_{2}^{\T b}a_{2}^{\T b}(\T c)
\label{eig:condition_energy}
\end{equation}
is satisfied.

The relation between the conditions \eqref{eig:condition_RuLaz} and \eqref{eig:condition_energy} depends on the signs of $\epsilon_{2}^{\T b}a_{2}^{\T b}(\T h)$ and $\epsilon_{2}^{\T b}a_{2}^{\T b}(\T c)$. The sign of the coefficient $\epsilon_{2}^{\T b}a_{2}^{\T b}(\T i)$ is related to the behavior of the average energy in Eq.~\eqref{eig:energy_evolution}. When this coefficient is positive, the average energy $\aveTt{\epsilon}{\T b}{\T i}$ approaches the equilibrium value $\langle\epsilon\rangle_{\T b}$ from above, which is an ordinary relaxation. When the coefficient is negative, the average energy approaches the equilibrium from below, exhibiting an anomalous undershooting. Below, let us assume for simplicity that $\epsilon_{2}^{\T b}a_{2}^{\T b}(\T h)$ and $\epsilon_{2}^{\T b}a_{2}^{\T b}(\T c)$ have the same sign, which is always the case if $\T h$ and $\T c$ are close enough to each other. When the average energies do not undershoot, i.e., $\epsilon_{2}^{\T b}a_{2}^{\T b}(\T h),\epsilon_{2}^{\T b}a_{2}^{\T b}(\T c) > 0$, the condition in Eq.~\eqref{eig:condition_energy} reduces to $\vert a_{2}^{\T b}(\T h)\vert < \vert a_{2}^{\T b}(\T c)\vert$. When the average energies undershoot, i.e., $\epsilon_{2}^{\T b}a_{2}^{\T b}(\T h),\epsilon_{2}^{\T b}a_{2}^{\T b}(\T c) < 0$, the condition is equivalent to $\vert a_{2}^{\T b}(\T h)\vert > \vert a_{2}^{\T b}(\T c)\vert$. Thus, unless the average energy undershoots, the condition in Eq.~\eqref{eig:condition_RuLaz} is still valid for the definition of ensemble MPEs in terms of the average energy.

\subsection{Eigenmode analysis of microstate MPEs}

While Eqs.~\eqref{eig:condition_RuLaz} and \eqref{eig:condition_energy} provide concise sufficient conditions for ensemble MPEs in the asymptotic regime, they do not reveal the microscopic origin of the ensemble MPEs\@. To get insight into the microscopic origin from the eigenmode analysis, we can combine the eigenmode expansion with our results on microstate MPEs.

For the definition of ensemble MPE in terms of the average energy in Eq.~\eqref{eq:ensemble_def}, we can apply the eigenmode analysis to our Result 1. Using Eq.~\eqref{eig:expansion_asymptotic}, the asymptotic behavior of the average $\avet{\epsilon}{\T b}{\bm{x}}$ reads
\begin{equation}
\avet{\epsilon}{\T b}{\bm{x}} \simeq \langle\epsilon\rangle_{\T b}+\exp(\lambda_{2}^{\T b}t)\epsilon_{2}^{\T b}\psi_{2}^{\T b}(\bm{x}),
\end{equation}
and thus, the measure of microstate MPE $c_{t}^{\T b}(\bm{x},\bm{y})$ is 
\begin{equation}
c_{t}^{\T b}(\bm{x},\bm{y}) \simeq \exp(\lambda_{2}^{\T b}t)\epsilon_{2}^{\T b}[\psi_{2}^{\T b}(\bm{x})-\psi_{2}^{\T b}(\bm{y})][\epsilon(\bm{x})-\epsilon(\bm{y})].
\label{eig:correlation}
\end{equation}
As discussed in the main text, our Result 1 implies that a pair of microstates $(\bm x,\bm y)$ has a dominant contribution to the ensemble MPE if $c_t^{\T b}(\bm x,\bm y)$ takes a large negative value. 
Therefore, one can numerically evaluate $\epsilon_{2}^{\T b}[\psi_{2}^{\T b}(\bm{x})-\psi_{2}^{\T b}(\bm{y})][\epsilon(\bm{x})-\epsilon(\bm{y})]$ using the second eigenfunctions to identify the major microscopic origin of ensemble MPEs in the long-time regime under the assumption $\lambda_{1}^{\T b} > \lambda_{2}^{\T b} > \lambda_{3}^{\T b}$.

For the definition of ensemble MPEs using the KL divergence in Eq.~\eqref{ent:ensemble_def}, we can similarly combine the eigenmode analysis with our Result S2 in Sec.~\ref{sec:entropic}\@. We first expand the nonequilibrium free energy as
\begin{equation}
f_{t}^{\T b}(\bm{x};\T i) \simeq  F_{\T b}+\kB\T b\exp(\lambda_{2}^{\T b}t)a_{2}^{\T b}(\T i)\frac{\phi_{2}^{\T b}(\bm{x})}{\piT{\T b}(\bm{x})},
\end{equation}
where we use Eq.~\eqref{eig:expansion_asymptotic} and $\ln(x+\delta) \simeq \ln(x) + (\delta/x)$ for $\delta \ll x$. Averaging this expression over $P_{t}^{\T b}(\bm{x}\vert\bm{y})$ and using Eq.~\eqref{eig:expansion_asymptotic} again, the average nonequilibrium free energy starting from a single state $\bm{y}$ is
\begin{equation}
\bar{f}_{t}^{\T b}(\bm{y};\T i) \simeq  F_{\T b}+\kB\T bA_2^{\T b}\exp(2\lambda_{2}^{\T b}t)a_{2}^{\T b}(\T i)\psi_{2}^{\T b}(\bm{y}),
\end{equation}
where we use $\intX x \,\phi_2^{\T b}(\bm x)=0$, and we define a positive constant $A_2^{\T b} \coloneqq \intX x\,\bigl[\phi_{2}^{\T b}(\bm{x})\bigr]^{2}/\piT{\T b}(\bm{x})$. The constant $A_2^{\T b}$ is typically unity (e.g., Sec.~7, Chapter V of Ref.~\cite{kampen2007stochastic}), but this property is not needed below. The measure of the microstate MPE appearing in our Result S2, $\tilde{c}_{t}^{\T b}(\bm{x},\bm{y};T_{\largestar})$ in Eq.~\eqref{ent:def_microstate}, is expanded as 
\begin{align}
\tilde{c}_{t}^{\T b}(\bm{x},\bm{y};T_{\largestar}) &  \simeq \left(1-\frac{\T b}{T_{\largestar}}\right)\kB\T bA_2^{\T b}\exp(2\lambda_{2}^{\T b}t)\nonumber \\
 & \quad\times a_{2}^{\T b}(T_{\largestar})[\psi_{2}^{\T b}(\bm{x})-\psi_{2}^{\T b}(\bm{y})][\epsilon(\bm{x})-\epsilon(\bm{y})],
\end{align}
where we used Eq.~\eqref{ent:f_is_e1}. From Result S2, a pair of states $(\bm x,\bm y)$ with a large negative $\tilde{c}_{t}^{\T b}(\bm{x},\bm{y}; T_{\largestar})$ constitutes a major source of the ensemble MPE defined with the KL divergence.
Thus, by numerically calculating the quantity $a_{2}^{\T b}(T_{\largestar})[\psi_{2}^{\T b}(\bm{x})-\psi_{2}^{\T b}(\bm{y})][\epsilon(\bm{x})-\epsilon(\bm{y})]$ from the second eigenfunctions, we can identify the primary source of ensemble MPEs in the asymptotic regime under the assumption $\lambda_{1}^{\T b} > \lambda_{2}^{\T b} > \lambda_{3}^{\T b}$.

\section{Proof of Result 2 
\label{sec:result2}}

\subsection{Preliminaries 
\label{subsec:apdx_absence_general}}

We prove Result 2 using classical results from the theory of stochastic interacting systems~\cite{Liggett2005InteractingParticle,Levin2017MarkovChains}. We introduce three definitions based on Chapter II of Ref.~\cite{Liggett2005InteractingParticle} and Chapter 22 of Ref.~\cite{Levin2017MarkovChains}. We fix a partial ordering $\preceq$ between states $\bm{x},\bm{y} \in \mathcal{X}$. We say a real-valued function $a(\bm{x})$ is \textit{monotone} if $a(\bm{x}) \leq  a(\bm{y})$ for all $\bm{x}\preceq\bm{y}$. We say that a Markov process is \textit{monotone} if, for any monotone function $a(\bm{x})$, $\avet a{\T b}{\bm{x}}$ is a monotone function for all $t\geq0$. We say that a probability distribution $q(\bm{x})$ has \textit{positive correlations} if 
\begin{align}
\sum_{x}a(\bm{x})b(\bm{x})q(\bm{x})-\biggl[\sum_{x}a(\bm{x})q(\bm{x})\biggr]\biggl[\sum_{x}b(\bm{x})q(\bm{x})\biggr] & \geq0
\label{res2:positive_correlation}
\end{align}
for all monotone functions $a(\bm{x})$ and $b(\bm{x})$.

We use our notation from the main text for the proof. We always use $R_{\T b}(\bm{x}\vert\bm{y})$ as a transition rate, $R_{\T b}^{\sharp}(\bm{u},\bm{v}\vert\bm{x},\bm{y})$ as a monotone coupling, and $P_{t}^{\T b}(\bm{x}\vert\bm{y})$ as the conditional probability generated by $R_{\T b}(\bm{x}\vert\bm{y})$ through the master equation in Eq.~\eqref{eq:master_equation}. As in the main text, we always assume that $P_{t}^{\T b}(\bm{x}\vert\bm{y})$ converges to $\piT{\T b}(\bm{x})$ as $t \to \infty$ from any $\bm{y} \in \mathcal{X}$. We use the set $\mathcal{K} \coloneqq \{(\bm{x},\bm{y}) \mid \bm{x}\preceq\bm{y}\}$.

The proof of Result 2 proceeds in three steps. In Sec.~\ref{subsec:apdx_monotonicity}, we introduce a theorem to prove the monotonicity of a Markov process. In Sec.~\ref{subsec:apdx_pos_corr}, we introduce another theorem to show that a probability distribution has positive correlations. Finally, we combine these theorems with the assumptions of Result 2 to prove the no-Mpemba theorem in Sec.~\ref{subsec:apdx_result2}. Note that the first and second steps are reviews of known results, but we provide proof of them for convenience.

\subsection{Monotonicity of the process
\label{subsec:apdx_monotonicity}}

We introduce a theorem to prove that a Markov process is monotone.

\paragraph{Theorem S4 (Proposition 22.7 of Ref.~\cite{Levin2017MarkovChains})}
If there exists a monotone coupling $R_{\T b}^{\sharp}(\doubleBullet\vert\doubleBullet)$ of $R_{\T b}(\Bullet\vert\Bullet)$, the Markov process generated by $R_{\T b}(\Bullet\vert\Bullet)$ is monotone.

\paragraph{Proof of Theorem S4}
Let $a(\bm{x})$ be any monotone function. We define three functions on $\mathcal{X}\times\mathcal{X}$, $a^{(1)}(\bm{u},\bm{v}) \coloneqq  a(\bm{u})$, $a^{(2)}(\bm{u},\bm{v}) \coloneqq  a(\bm{v})$, and $a^{(3)}(\bm{u},\bm{v}) \coloneqq  a(\bm{v})-a(\bm{u})$. We readily have $a^{(3)}(\bm{u},\bm{v})\geq0$ for $(\bm{u},\bm{v}) \in \mathcal{K}$ due to the definition of a monotone function. For $i = 1,2,3$, we use $\overline{a^{(i)}}_{\,t}^{\T b}(\bm{x},\bm{y})$ to denote the average of $a^{(i)}(\doubleBullet)$ at time $t$ over the coupled process starting with a fixed initial state $(\bm{x},\bm{y})$ at time $0$. For any $(\bm{x},\bm{y}) \in \mathcal{K}$ and any $t\ge0$, we obtain
\begin{align}
\avet a{\T b}{\bm{y}}-\avet a{\T b}{\bm{x}} &  = \overline{a^{(2)}}_{\,t}^{\T b}(\bm{x},\bm{y})-\overline{a^{(1)}}_{\,t}^{\T b}(\bm{x},\bm{y})\nonumber \\
 &  = \overline{a^{(3)}}_{\,t}^{\T b}(\bm{x},\bm{y})\geq0.
\label{res2:proof_monotone}
\end{align}
Here, the first equality holds because each system copy in the monotone coupling has the same marginal transition rates as the original process, and thus, the statistics about each copy are the same as the statistics over the original process. The second equality follows from $a^{(3)}(\bm{u},\bm{v}) = a^{(2)}(\bm{u},\bm{v})-a^{(1)}(\bm{u},\bm{v})$. The last inequality holds because the coupled process starting with $(\bm{x},\bm{y}) \in \mathcal{K}$ stays in $\mathcal{K}$, over which $a^{(3)}$ is nonnegative. 

In summary, we have proved that for any monotone function $a(\bm{x})$ and for any $\bm{x}\preceq\bm{y}$, $\avet a{\T b}{\bm{x}} \leq \avet a{\T b}{\bm{y}}$ for all $t\geq0$. This statement is identical to the definition of a monotone process. \qed

If the partial order $\preceq$ happens to be a total order, Result 2 directly follows from this Theorem S4. Indeed, if $\preceq$ is a total order, any pair of states $(\bm{x},\bm{y})$ are either $\bm{x}\preceq\bm{y}$ or $\bm{x}\succeq\bm{y}$. If $\bm{x}\preceq\bm{y}$, the assumption (a) of Result 2 implies $\epsilon(\bm{x}) \leq \epsilon(\bm{y})$, and combining assumption (c) of Result 2 and Theorem S4 gives $\avet{\epsilon}{\T b}{\bm{x}} \leq \avet{\epsilon}{\T b}{\bm{y}}$. Therefore, we have $c_{t}^{\T b}(\bm{x},\bm{y})\geq0$. If $\bm{x}\succeq\bm{y}$, we can similarly show $c_{t}^{\T b}(\bm{x},\bm{y})\geq0$ by exchanging the roles of $\bm{x}$ and $\bm{y}$. Thus, $c_{t}^{\T b}(\bm{x},\bm{y})\geq0$ for all pairs of states, i.e., the microstate MPE is absent. This implies $C_{t}^{\T b}(T_{\largestar})\geq0$ for any $T_{\largestar}$, namely the absence of ensemble MPEs. 

For a general partial ordering $\preceq$, Theorem S4 alone is not sufficient to prove Result 2. We introduce another element of the proof below.

\subsection{Positive correlations 
\label{subsec:apdx_pos_corr}}

We introduce a classical theorem to prove that a probability distribution has positive correlations.

\paragraph{Lemma S5 (Main theorem of Ref.~\cite{Harris1977CorrelationInequality})}
Suppose that the Markov process generated by $R_{\T b}(\Bullet\vert\Bullet)$ is monotone. Then, the following two statements are equivalent: \textit{(a) The transition rate $R_{\T b}(\bm{x}\vert\bm{y})$ is nonzero only if $\bm{x}$ and $\bm{y}$ are comparable (i.e., $\bm{x}\preceq\bm{y}$ or $\bm{y}\preceq\bm{x}$); (b) For any probability distribution $q(\Bullet)$ that has positive correlations, the probability distribution $\sum_{\bm{y}}P_{t}^{\T b}(\Bullet\vert\bm{y})q(\bm{y})$ has positive correlations. }

\vspace{\sectionsep}

We omit the proof of this Lemma. The proof in the original paper~\cite{Harris1977CorrelationInequality} is rather complicated. A simpler proof is found in Theorem 2.14, Chapter II of Ref.~\cite{Liggett2005InteractingParticle}. We then use this Lemma to deduce that the Gibbs distributions have positive correlations.

\paragraph{Theorem S6~(Theorem 22.16 of Ref.~\cite{Levin2017MarkovChains})}
Suppose that \textit{(a) The Markov process generated by $R_{\T b}(\Bullet\vert\Bullet)$ is monotone, and (b) $R_{\T b}(\bm{x}\vert\bm{y})$ is nonzero only if $\bm{x}$ and $\bm{y}$ are comparable.} Then $\piT{\T b}(\Bullet)$ has positive correlations for any $t\geq0$.

\paragraph{Proof of Theorem S6}
Under the assumptions of Theorem S6, we can use Lemma S5 to conclude that $\sum_{\bm{y}}P_{t}^{\T b}(\Bullet\vert\bm{y})q(\bm{y})$ has positive correlations for any $q(\Bullet)$ that has positive correlations. 

Take an arbitrary state $\bm{z} \in \mathcal{X}$, and take the distribution concentrated on $\bm{z}$, $q(\bm{x}) = \delta_{\bm{x}\bm{z}}$. This distribution has positive correlations because
\begin{align}
 & \sum_{x}a(\bm{x})b(\bm{x})\delta_{\bm{x}\bm{z}}-\biggl[\sum_{x}a(\bm{x})\delta_{\bm{x}\bm{z}}\biggr]\biggl[\sum_{x}b(\bm{x})\delta_{\bm{x}\bm{z}}\biggr]\nonumber \\
 & \hspace{8em} = a(\bm{z})b(\bm{z})-a(\bm{z})b(\bm{z}) = 0.
\label{res2:proof_positive_correlation}
\end{align}
Then, $\sum_{\bm{y}}P_{t}^{\T b}(\Bullet\vert\bm{y})\delta_{\bm{y}\bm{z}} = P_{t}^{\T b}(\Bullet\vert\bm{z})$ has positive correlations. Sending to the limit $t \to \infty$, $P_{t}^{\T b}(\Bullet\vert\bm{z})$ converges to $\piT{\T b}(\Bullet)$, and hence, $\piT{\T b}(\Bullet)$ has positive correlations. \qed

\subsection{Proof of Result 2
\label{subsec:apdx_result2}}

We combine Theorems S4 and S6 to prove Result 2 of the main text.

\paragraph{Proof of Result 2}
Combining assumption (c) of Result 2 with Theorem S4, the Markov process generated by $R_{\T b}(\Bullet\vert\Bullet)$ is monotone for all $\T b \in [T_{1},T_{2}]$. Assumption (a) of Result 2 says that the energy $\epsilon(\bm{x})$ is a monotone function of $\bm{x}$. Thus, from the definition of a monotone process, $\avet{\epsilon}{\T b}{\bm{x}}$ is a monotone function of $\bm{x}$ for all $\T b \in [T_{1},T_{2}]$ and all $t\geq 0$.

Next, by combining assumption (b) of Result 2 and the monotonicity of $R_{\T b}(\Bullet\vert\Bullet)$ for all $\T b \in [T_{1}, T_{2}]$, we can use Theorem S6 to deduce that $\piT{\T b}(\Bullet)$ has positive correlations for all $\T b \in [T_{1},T_{2}]$. By simply replacing the symbol, $\piT{T_{\largestar}}(\Bullet)$ has positive correlations for all $T_{\largestar} \in [T_{1},T_{2}]$.

We use the definition of having positive correlations to obtain 
\begin{align}
 & \sum_{\bm{x}}\avet{\epsilon}{\T b}{\bm{x}}\epsilon(\bm{x})\piT{T_{\largestar}}(\bm{x})\nonumber \\
 & -\biggl[\sum_{\bm{x}}\avet{\epsilon}{\T b}{\bm{x}}\piT{T_{\largestar}}(\bm{x})\biggr]\biggl[\sum_{\bm{x}}\epsilon(\bm{x})\piT{T_{\largestar}}(\bm{x})\biggr]\geq0
\label{res2:proof_positiveC}
\end{align}
for any $\T b,T_{\largestar} \in [T_{1},T_{2}]$. The left-hand side of Eq.~\eqref{res2:proof_positiveC} is equal to $\frac{1}{2}C_{t}^{\T b}(T_{\largestar})$, which is easily shown using the second line of Eq.~\eqref{pr:rearrange}. Thus, $C_{t}^{\T b}(T_{\largestar})\geq0$ for any $\T b,T_{\largestar} \in [T_{1},T_{2}]$ and any $t\geq 0$.

Finally, we prove the absence of ensemble MPE for any $\T b,\T c,\T h \in [T_{1},T_{2}]$ by contradiction. Suppose the ensemble MPE is present at time $t$ for a set of temperatures $\T b,\T c,\T h \in [T_{1},T_{2}]$ with $\T b < \T c < \T h$. Then, Result 1 says that there exists a temperature $T_{\largestar}$ such that $\T c < T_{\largestar} < \T h$ and $C_{t}^{\T b}(T_{\largestar}) < 0$. However, this contradicts the conclusion in the previous paragraph. Therefore, the ensemble MPE is absent for any $\T b,\T c,\T h \in [T_{1},T_{2}]$.
\qed

\section{Applications of the no-Mpemba theorem 
\label{sec:result2_application}}

\subsection{Ferromagnetic Ising model 
\label{subsec:apdx_absence_Ising}}

We provide the details of the application of the no-Mpemba theorem to the ferromagnetic Ising model under strong magnetic fields. Before doing so, we check that the transition rates of the Ising model are irreducible and have $\piT{\T b}(\Bullet)$ as the stationary distribution, which we assumed in the main text. The irreducibility is confirmed by considering that, starting from any spin configuration, we can reach any other configuration by flipping spins one by one. The model has the stationary distribution  $\piT{\T b}(\Bullet)$ because it satisfies the detailed-balance condition,
\begin{align}
\frac{R_{\T b}(\bm{x}^{i}\vert\bm{x})}{R_{\T b}(\bm{x}\vert\bm{x}^{i})} &  = \frac{r_{i}^{\T b}(\epsilon(\bm{x}^{i})-\epsilon(\bm{x}))}{r_{i}^{\T b}(\epsilon(\bm{x})-\epsilon(\bm{x}^{i}))}\nonumber \\
 &  = \exp\left[\frac{\epsilon(\bm{x})-\epsilon(\bm{x}^{i})}{\kB\T b}\right] = \frac{\piT{\T b}(\bm{x}^{i})}{\piT{\T b}(\bm{x})},
\end{align}
where we used the detailed-balance condition for $r_{i}^{\T b}(\Delta\epsilon)$ in the main text.

We check the conditions (a)--(c) of Result 2 in order. To check condition (a) of Result 2, let $\bm{x}$ and $\bm{y}$ be two states that satisfy $\bm{x}\preceq\bm{y}$. For every spin $i$ such that $x_{i} \neq  y_{i}$, it is guaranteed that $-1 = x_{i} < y_{i} = +1$. Therefore, $\bm{y}$ is obtained from $\bm{x}$ by flipping those spins from $-1$ to $+1$. By flipping a single spin $i$ from $-1$ to $+1$ of an arbitrary state $\bm{z}$ that has $z_i=-1$, the energy changes by
\begin{align}
\epsilon(\bm{z}^{i})-\epsilon(\bm{z}) &  = H_{i}[1-(-1)]-\sum_{j}J_{ij}z_{j}[1-(-1)]\nonumber \\[-0.5em]
 &  = 2H_{i}-2\sum_{j}J_{ij}z_{j}.
\label{mo:energy_diff}
\end{align}
This energy change is positive because
\begin{equation}
H_{i}-\sum_{j}J_{ij}z_{j} \geq  H_{i}-\sum_{j}J_{ij}\vert z_{j}\vert = H_{i}-\sum_{j}J_{ij}\geq0,
\end{equation}
where the first equality uses $J_{ij}\geq0$, and the last inequality is the assumption of a strong magnetic field in the main text. Since $\bm{y}$ is obtained from $\bm{x}$ by consecutively flipping spins from $-1$ to $+1$, the total energy change is positive, i.e., $\epsilon(\bm{x}) \leq \epsilon(\bm{y})$. This confirms condition (a) of Result 2. 

Condition (b) of Result 2 can be checked by noting that, for any state $\bm{x}$ and for every spin $i$, $\bm{x}\preceq\bm{x}^{i}$ if $x_{i} = -1$ and $\bm{x}\succeq\bm{x}^{i}$ if $x_{i} = +1$. The transitions are only between $\bm{x}$ and $\bm{x}^{i}$, thus between comparable states.

To check condition (c) of Result 2, we use a monotone coupling found in Ref.~\cite{Liggett2005InteractingParticle} (see Theorems 1.5 and 2.2 of Chapter III therein). For each coupled state $(\bm{x},\bm{y}) \in \mathcal{X}\times\mathcal{X}$ and for each spin $i$, we set the following transition rates: If $x_{i} = y_{i}$, 
\begin{align}
R_{\T b}^{\sharp}(\bm{x}^{i},\bm{y}^{i}\vert\bm{x},\bm{y}) &  = \min\{R_{\T b}(\bm{x}^{i}\vert\bm{x}),\,R_{\T b}(\bm{y}^{i}\vert\bm{y})\},\nonumber \\
R_{\T b}^{\sharp}(\bm{x}^{i},\bm{y}\vert\bm{x},\bm{y}) &  = R_{\T b}(\bm{x}^{i}\vert\bm{x})-\min\{R_{\T b}(\bm{x}^{i}\vert\bm{x}),\,R_{\T b}(\bm{y}^{i}\vert\bm{y})\},\nonumber \\
R_{\T b}^{\sharp}(\bm{x},\bm{y}^{i}\vert\bm{x},\bm{y}) &  = R_{\T b}(\bm{y}^{i}\vert\bm{y})-\min\{R_{\T b}(\bm{x}^{i}\vert\bm{x}),\,R_{\T b}(\bm{y}^{i}\vert\bm{y})\},
\label{mo:ising_rate_coupled}
\end{align}
and if $x_{i} \neq  y_{i}$, 
\begin{equation}
\begin{aligned}
R_{\T b}^{\sharp}(\bm{x}^{i},\bm{y}\vert\bm{x},\bm{y}) &  = R_{\T b}(\bm{x}^{i}\vert\bm{x}),\\
R_{\T b}^{\sharp}(\bm{x},\bm{y}^{i}\vert\bm{x},\bm{y}) &  = R_{\T b}(\bm{y}^{i}\vert\bm{y}).
\end{aligned}
\label{mo:ising_rate_decoupled}
\end{equation}
We set any other $R_{\T b}^{\sharp}(\bm{u},\bm{v}\vert\bm{x},\bm{y})$ with $(\bm{x},\bm{y}) \neq (\bm{u},\bm{v})$ to zero. With these coupled transition rates, the marginal transition rates of the first copy are the same as the original transition rates because 
\begin{align}
\sum_{\bm{v}}R_{\T b}^{\sharp}(\bm{x}^{i},\bm{v}\vert\bm{x},\bm{y}) &= R_{\T b}^{\sharp}(\bm{x}^{i},\bm{y}\vert\bm{x},\bm{y})+R_{\T b}^{\sharp}(\bm{x}^{i},\bm{y}^{i}\vert\bm{x},\bm{y})\nonumber \\[-0.5em]
 &  = R_{\T b}(\bm{x}^{i}\vert\bm{x})
\end{align}
for every $\bm{x},\bm{y}$ and $i$. The second copy satisfies a similar property. Moreover, the coupled process satisfies $R_{\T b}^{\sharp}(\bm{u},\bm{v}\vert\bm{x},\bm{y}) = 0$ if $(\bm{x},\bm{y}) \in \mathcal{K}$ and $(\bm{u},\bm{v})\notin\mathcal{K}$, as proven in the following.

To prove this property, we first show that the transition rates of the original model satisfy the so-called \textit{attractive} property (Sec.~2, Chap.~III of Ref.~\cite{Liggett2005InteractingParticle}). The attractive property means that, for every $\bm{x}$ and $\bm{y}$ such that $\bm{x}\preceq\bm{y}$,
\begin{align}
R_{\T b}(\bm{x}^{i}\vert\bm{x}) &  \leq  R_{\T b}(\bm{y}^{i}\vert\bm{y}) &  & \text{for all \ensuremath{i} such that \ensuremath{x_{i} = y_{i} = -1}},\nonumber \\
R_{\T b}(\bm{x}^{i}\vert\bm{x}) &  \geq  R_{\T b}(\bm{y}^{i}\vert\bm{y}) &  & \text{for all \ensuremath{i} such that \ensuremath{x_{i} = y_{i} = +1}}.
\label{mo:attractive}
\end{align}
This condition says that if $\bm{x}\preceq\bm{y}$, the state $\bm{y}$ has a higher flip rate from $-1$ to $+1$ and a lower flip rate from $+1$ to $-1$ than $\bm{x}$. We check this attractive property for each $\bm{x}\preceq \bm{y}$ and $i$ as follows. If $x_{i} = y_{i} = -1$, Eq.~\eqref{mo:energy_diff} implies
\begin{equation}
[\epsilon(\bm{y}^{i})-\epsilon(\bm{y})]-[\epsilon(\bm{x}^{i})-\epsilon(\bm{x})] = -2\sum_{j}J_{ij}(y_{j}-x_{j})\leq0
\label{mo:attractive_energy}
\end{equation}
because the coupling is ferromagnetic $J_{ij}\geq0$, and $\bm{x}\preceq\bm{y}$ implies $x_{j} \leq  y_{j}$. Since the transition rate is $R_{\T b}(\bm{x}^{i}\vert\bm{x}) = r_{i}^{\T b}(\epsilon(\bm{x}^{i})-\epsilon(\bm{x}))$, and $r_{i}^{\T b}(\Delta\epsilon)$ is a decreasing function of $\Delta\epsilon$, Eq.~\eqref{mo:attractive_energy} implies $R_{\T b}(\bm{y}^{i}\vert\bm{y}) \geq  R_{\T b}(\bm{x}^{i}\vert\bm{x})$. This proves the first equation in Eq.~\eqref{mo:attractive}. If $x_{i} = y_{i} = +1$, we can similarly show the second equation in Eq.~\eqref{mo:attractive}. 

We use this attractive property to show that, for every $(\bm{x},\bm{y}) \in \mathcal{K}$ and for each $i$, the destinations of the transitions determined by Eqs.~\eqref{mo:ising_rate_coupled} and \eqref{mo:ising_rate_decoupled} are in $\mathcal{K}$. If $x_{i} = y_{i} = -1$, the attractive property implies $\min\{R_{\T b}(\bm{x}^{i}\vert\bm{x}),\,R_{\T b}(\bm{y}^{i}\vert\bm{y})\} = R_{\T b}(\bm{x}^{i}\vert\bm{x})$, thus $R_{\T b}^{\sharp}(\bm{x}^{i},\bm{y}\vert\bm{x},\bm{y}) = 0$ in Eq.~\eqref{mo:ising_rate_coupled}. Therefore, there are only two possible transitions, $(\bm{x},\bm{y}) \to (\bm{x}^{i},\bm{y}^{i})$ and $(\bm{x},\bm{y}) \to (\bm{x},\bm{y}^{i})$. The destinations of these transitions satisfy $(\bm{x}^{i},\bm{y}^{i}) \in \mathcal{K}$ and $(\bm{x},\bm{y}^{i}) \in \mathcal{K}$. Similarly, if $x_{i} = y_{i} = +1$, the possible transitions are $(\bm{x},\bm{y}) \to (\bm{x}^{i},\bm{y}^{i})$ and $(\bm{x},\bm{y}) \to (\bm{x}^{i},\bm{y})$, and the destinations satisfy $(\bm{x}^{i},\bm{y}^{i}) \in \mathcal{K}$ and $(\bm{x}^{i},\bm{y}) \in \mathcal{K}$. If $-1 = x_{i} < y_{i} = +1$, the possible transitions are $(\bm{x},\bm{y}) \to (\bm{x}^{i},\bm{y})$ and $(\bm{x},\bm{y}) \to (\bm{x},\bm{y}^{i})$, whose destinations are in $\mathcal{K}$. This completes the verification of condition (c) of Result 2.

Note that the assumption about the transition rates, $R_{\T b}(\bm{x}^{i}\vert\bm{x}) = r_{i}^{\T b}(\epsilon(\bm{x}^{i})-\epsilon(\bm{x}))$ with $r_{i}^{\T b}(\Bullet)$ decreasing, is sufficient but not necessary for the attractive property. Since we did not directly use the form $R_{\T b}(\bm{x}^{i}\vert\bm{x}) = r_{i}^{\T b}(\epsilon(\bm{x}^{i})-\epsilon(\bm{x}))$ to verify condition (c) of Result 2, we can apply our Result 2 to a slightly broader class of systems by removing the assumption $R_{\T b}(\bm{x}^{i}\vert\bm{x}) = r_{i}^{\T b}(\epsilon(\bm{x}^{i})-\epsilon(\bm{x}))$ and directly assuming the attractive property and the detailed balance condition.

\subsection{Exclusion process 
\label{subsec:apdx_absence_exclusion}}

We discuss the details of the application of Result 2 to the one-dimensional exclusion process. We first define some notation. For a state $\bm{x}$ and a site $k \in \{1,\dots,L\}$, we say $\bm{x}^{k+}$ is valid if $1 \leq  k \leq  L-1$, site $k$ is occupied, and site $k+1$ is empty in $\bm{x}$. Under these conditions, we define $\bm{x}^{k+}$ as the state obtained from $\bm{x}$ by moving the particle from site $k$ to $k+1$. Similarly, we say $\bm{x}^{k-}$ is valid if $2 \leq  k \leq  L$, site $k$ is occupied, and site $k-1$ is empty, and under these conditions, we define $\bm{x}^{k-}$ similarly. The transition rates are $R_{\T b}(\bm{x}^{k+}\vert\bm{x}) = r_{k+1\vert k}^{\T b}$ if $\bm{x}^{k+}$ is valid, $R_{\T b}(\bm{x}^{k-}\vert\bm{x}) = r_{k-1\vert k}^{\T b}$ if $\bm{x}^{k-}$ is valid, and zero otherwise.

Before applying Result 2, we verify that the exclusion process is irreducible and has $\piT{\T b}(\Bullet)$ as the stationary distribution. The process is irreducible because, starting with any configuration, the process can reach any other configuration by moving particles one at a time. The process has the stationary distribution $\piT{\T b}(\Bullet)$ because the detailed-balance condition
\begin{equation}
\frac{R_{\T b}(\bm{x}^{k+}\vert\bm{x})}{R_{\T b}(\bm{x}\vert\bm{x}^{k+})} = \frac{r_{k+1\vert k}^{\T b}}{r_{k\vert k+1}^{\T b}} = \exp\left[\frac{e_{k}-e_{k+1}}{\kB\T b}\right] = \frac{\piT{\T b}(\bm{x}^{k+})}{\piT{\T b}(\bm{x})}
\end{equation}
is satisfied for any $\bm{x}$ whose $\bm{x}^{k+}$ is valid. 

We check conditions (a)--(c) of Result 2 one by one. Condition (a) of Result 2 is satisfied because, for any $\bm{x}\preceq\bm{y}$,
\begin{equation}
\epsilon(\bm{y})-\epsilon(\bm{x}) = \sum_{i = 1}^{N}(e_{y_{i}}-e_{x_{i}})\geq0.
\end{equation}
The last inequality holds because $\bm{x}\preceq\bm{y}$ implies $x_{i} \leq  y_{i}$, which in turn implies $e_{x_{i}} \leq  e_{y_{i}}$ for a monotone potential.

Condition (b) of Result 2 is verified by noting that $\bm{x}\preceq\bm{x}^{k+}$ and $\bm{x}\succeq\bm{x}^{k-}$. The transitions occur only between $\bm{x}$ and $\bm{x}^{k+}$ and between $\bm{x}$ and $\bm{x}^{k-}$. Thus, they occur only between comparable states.

\begin{figure}
\includegraphics[width = 1\columnwidth]{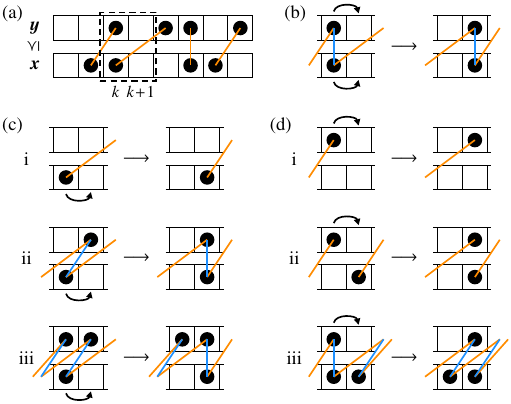}

\caption{Proof that the monotone coupling stays in the set $\mathcal{K}$ after performing a jump in Eq.~\eqref{mo:exclusion_couple1} from $(\bm{x},\bm{y}) \in \mathcal{K}$. The jumps in Eq.~\eqref{mo:exclusion_couple2} are treated similarly. (a) For $(\bm{x},\bm{y}) \in \mathcal{K}$, we draw the two states in a diagram, where \raisebox{-0.1em}{\scalebox{1.4}{$\square$}} is a site, and $\CIRCLE$ represents a particle. We connect the $i$th particle in $\bm{x}$ and the $i$th particle in $\bm{y}$ with a bond (orange) for $i = 1,\dots,N$. The property $\bm{x}\preceq\bm{y}$ is graphically represented as all the bonds having the directions $\bm{\vert}$ or $\diagup$, not $\diagdown$. Below, we focus on two adjacent sites, $k$ and $k+1$ (encircled by a dashed line). (b) Case for the first equation in Eq.~\eqref{mo:exclusion_couple1}, i.e., when $\bm{x}^{k+}$ and $\bm{y}^{k+}$ are both valid. There are two possible patterns of the bonds before the jump, shown in orange and blue in the left panel of (b). Blue indicates that the two particles on the $k$th sites are connected with each other, while orange indicates that the two particles are connected with particles outside the focused region. The right panel of (b) shows the patterns of the bonds after the jump $(\bm{x},\bm{y}) \to (\bm{x}^{k+},\bm{y}^{k+})$. All the bonds have the directions $\bm{\vert}$ or $\diagup$, meaning that $(\bm{x}^{k+},\bm{y}^{k+}) \in \mathcal{K}$. (c) Cases for the second equation in Eq.~\eqref{mo:exclusion_couple1}, i.e., $\bm{x}^{k+}$ is valid and $\bm{y}^{k+}$ is not. There are three possible configurations of the particles before the jump (i, ii, and iii), and each of them has one or two possible patterns of the bonds (orange and blue). In any of these cases, the jump $(\bm{x},\bm{y}) \to (\bm{x}^{k+},\bm{y})$ will keep the directions of all the bonds $\bm{\vert}$ or $\diagup$, meaning that the coupled system stays in $\mathcal{K}$. (d) Cases for the third equation in Eq.~\eqref{mo:exclusion_couple1}. In any case, the system remains in $\mathcal{K}$ after the jump $(\bm{x},\bm{y}) \to (\bm{x},\bm{y}^{k+})$.
\label{fig:proof_exclusion}}
\end{figure}

Condition (c) of Result 2 is satisfied with the following monotone coupling (Sec.~23.1 of Ref.~\cite{Levin2017MarkovChains}). For every coupled state $(\bm{x},\bm{y}) \in \mathcal{X}\times\mathcal{X}$ and for each site $i$, we set the following transition rates:
\begin{align}
R_{\T b}^{\sharp}(\bm{x}^{k+},\bm{y}^{k+}\vert\bm{x},\bm{y}) &  = r_{k+1\vert k}^{\T b} &  & \text{if \ensuremath{\bm{x}^{k+}} and \ensuremath{\bm{y}^{k+}} are both valid,}\nonumber \\
R_{\T b}^{\sharp}(\bm{x}^{k+},\bm{y}\vert\bm{x},\bm{y}) &  = r_{k+1\vert k}^{\T b} &  & \text{if \ensuremath{\bm{x}^{k+}} is valid and \ensuremath{\bm{y}^{k+}} is not,}\nonumber \\
R_{\T b}^{\sharp}(\bm{x},\bm{y}^{k+}\vert\bm{x},\bm{y}) &  = r_{k+1\vert k}^{\T b} &  & \text{if \ensuremath{\bm{y}^{k+}} is valid and \ensuremath{\bm{x}^{k+}} is not,}
\label{mo:exclusion_couple1}
\end{align}
and
\begin{align}
R_{\T b}^{\sharp}(\bm{x}^{k-},\bm{y}^{k-}\vert\bm{x},\bm{y}) &  = r_{k-1\vert k}^{\T b} &  & \text{if \ensuremath{\bm{x}^{k-}} and \ensuremath{\bm{y}^{k-}} are both valid,}\nonumber \\
R_{\T b}^{\sharp}(\bm{x}^{k-},\bm{y}\vert\bm{x},\bm{y}) &  = r_{k-1\vert k}^{\T b} &  & \text{if \ensuremath{\bm{x}^{k-}} is valid and \ensuremath{\bm{y}^{k-}} is not,}\nonumber \\
R_{\T b}^{\sharp}(\bm{x},\bm{y}^{k-}\vert\bm{x},\bm{y}) &  = r_{k-1\vert k}^{\T b} &  & \text{if \ensuremath{\bm{y}^{k-}} is valid and \ensuremath{\bm{x}^{k-}} is not.}
\label{mo:exclusion_couple2}
\end{align}
Namely, the particles at site $k$ of both copies move simultaneously if the move is valid for both copies, and otherwise, the particle at site $k$ moves alone. We set any other rate $R_{\T b}^{\sharp}(\bm{u},\bm{v}\vert\bm{x},\bm{y})$ with $(\bm{x},\bm{y}) \neq (\bm{u},\bm{v})$ to zero. The marginal transition rates of the first copy are the same as $R_{\T b}(\Bullet\vert\Bullet)$ because 
\begin{align}
\sum_{\bm{v}}R_{\T b}^{\sharp}(\bm{x}^{k+},\bm{v}\vert\bm{x},\bm{y}) &  = R_{\T b}^{\sharp}(\bm{x}^{k+},\bm{y}\vert\bm{x},\bm{y})+R_{\T b}^{\sharp}(\bm{x}^{k+},\bm{y}^{k+}\vert\bm{x},\bm{y})\nonumber \\[-0.5em]
 &  = r_{k+1\vert k}^{\T b} = R_{\T b}(\bm{x}^{k+}\vert\bm{x}),
\end{align}
and similarly for $\sum_{\bm{v}}R_{\T b}^{\sharp}(\bm{x}^{k-},\bm{v}\vert\bm{x},\bm{y})$. A similar property holds for the second copy. The transition rates satisfy $R_{\T b}^{\sharp}(\bm{u},\bm{v}\vert\bm{x},\bm{y}) = 0$ if $(\bm{x},\bm{y}) \in \mathcal{K}$ and $(\bm{u},\bm{v})\notin\mathcal{K}$. This property is confirmed by showing that all possible destinations of the transitions from $(\bm{x},\bm{y}) \in \mathcal{K}$ are in $\mathcal{K}$, which is checked by cases in Fig.~\ref{fig:proof_exclusion}.

\section{Supplemental numerical results 
\label{sec:numerics}}

\subsection{Single particle in a one-dimensional continuous space}

\begin{figure}[t]
\includegraphics[width = 1\columnwidth]{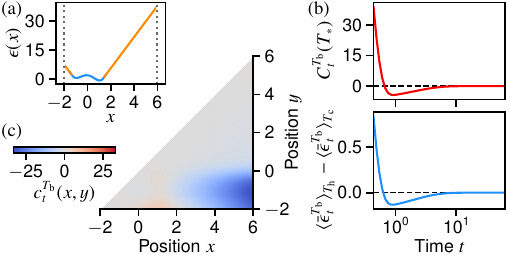}
\caption{Result 1 applied to a model of a Brownian particle in a one-dimensional double-well potential. (a) Potential function composed of a quartic potential (blue) and two linear potentials (orange). Reflective boundaries are at $x = -2,6$ (gray). (b) Existence of the ensemble MPEs for $\kB\T b = 1$, shown by the negativity of $C_{t}^{\T b}(T_{\largestar})$ for $\kB T_\largestar = 1000$ (upper panel) and negative values of the energy difference $\aveTt{\epsilon}{\T b}{\T h}-\aveTt{\epsilon}{\T b}{\T c}$ for $(\kB\T h,\kB\T c) = (2000,10)$ (lower panel). (c) The measure of microstate MPEs $c_{t}^{\T b}(x,y)$ at $t = 1.0$. Blue indicates negative values, i.e., the existence of microstate MPE. 
\label{fig:continuous}}
\end{figure}

We use our Result 1 to analyze ensemble MPEs in a single particle system in a one-dimensional continuous space. This class of systems has served as a prototype for studying MPEs~\cite{Kumar2020ExponentiallyFaster,Kumar2022AnomalousHeating,Lu2017NonequilibriumThermodynamics,Chtrite2021TheMetastableMpemba,Walker2021AnomalousThermal,Walker2022MpembaEffect,Schwarzendahl2022AnomalousCooling,Deguenther2022AnomalousRelaxation,Biswas2023MpembaEffect1,Biswas2023MpembaEffect2,Teza2023RelaxationShortcuts,Walker2023OptimalTransport}, and it has been experimentally realized using a colloidal particle~\cite{Kumar2020ExponentiallyFaster,Kumar2022AnomalousHeating}. 
To be concrete, we use the double-well potential used in the experiment in Ref.~\cite{Kumar2020ExponentiallyFaster}. The potential is composed of a quartic potential and two linear potentials continuated so that the potential and its derivative are continuous [Fig.~\ref{fig:continuous}(a)]. 
Letting $x$ denote the (properly non-dimensionalized) position variable, the potential is given by
\begin{equation}
\epsilon(x) = \begin{cases}
\epsilon_{\mathrm{quart}}(x_{\mathrm{L}})-8(x-x_{\mathrm{L}}) & (-2 \leq  x < x_{\mathrm{L}})\\
\epsilon_{\mathrm{quart}}(x) & (x_{\mathrm{L}} \leq  x \leq  x_{\mathrm{R}})\\
\epsilon_{\mathrm{quart}}(x_{\mathrm{R}})+8(x-x_{\mathrm{R}}) & (x_{\mathrm{R}} < x\leq6)
\end{cases}
\end{equation}
where $\epsilon_{\mathrm{quart}}(x)$ is a tilted quartic potential centered at $x = 0$,
\begin{equation}
\epsilon_{\mathrm{quart}}(x) = 2(x^{2}-1)^{2}-0.65x,
\end{equation}
and $x_{\mathrm{L}}$ and $x_{\mathrm{R}}$ are determined numerically by $(d\epsilon_{\mathrm{quart}}/dx)\vert_{x = x_{\mathrm{L}}} = -8$ and $(d\epsilon_{\mathrm{quart}}/dx)\vert_{x = x_{\mathrm{R}}} = 8$ so that the derivative of the potential is continuous. 
Here, we use the parameter values in Ref.~\cite{Kumar2020ExponentiallyFaster}, while the following results are not sensitive to the specific values of parameters. We set reflective boundaries at $x = -2$ and $6$. 
This asymmetry of the domain is essential for observing ensemble MPEs~\cite{Kumar2020ExponentiallyFaster}. 
The position $x$ obeys a Langevin equation, and hence, the conditional probability density $P_{t}^{\T b}(x\vert y)$ is determined by the Fokker--Planck equation:
\begin{equation}
\frac{\partial}{\partial t}P_{t}^{\T b}(x\vert y) = \frac{\partial}{\partial x}\left[P_{t}^{\T b}(x\vert y)\frac{\partial\epsilon}{\partial x}+\kB \T b\frac{\partial}{\partial x}P_{t}^{\T b}(x\vert y)\right]
\label{num:Fokker-Planck}
\end{equation}
with the initial condition $P_{0}^{\T b}(x\vert y) = \delta(x-y)$, where we set the mobility to unity.

This system exhibits the ensemble MPE, as shown in Fig.~\ref{fig:continuous}(b). To analyze its microscopic origin using our Result 1, we plot $c_{t}^{\T b}(x,y)$ for all pairs of states $(x,y)$ with $x \geq  y$ in Fig.~\ref{fig:continuous}(c). Almost all negative contributions appear in the region of $x > 2$ and $y < 0$, i.e., $x$ in the linear part of the right well and $y$ in the left well. This observation leads to the following mechanism of the ensemble MPE\@. Trajectories starting with the linear part of the right well relax quickly to the bottom of the right well, while trajectories starting with the left well remain in the left well for a long time. Since the bottom of the left well is higher than the bottom of the right well, this causes a crossing between $\avet{\epsilon}{\T b}x$ and $\bar{\epsilon}_{t}^{\T b}(y)$, i.e., the microstate MPE\@. This effect accumulates over almost all $(x,y)$ with $x > 2$ and $y < 0$, leading to a negative value of $C_{t}^{\T b}(T_{\largestar})$, which implies the ensemble MPE\@. In this way, our Result 1 can provide an intuitive explanation of the microscopic mechanism of ensemble MPEs.

\subsection{Numerical test of Result 2}

We numerically verify the no-Mpemba theorem applied to the ferromagnetic Ising models and the one-dimensional exclusion processes introduced in the main text. The no-Mpemba theorem establishes the absence of ensemble MPEs under a very general setup, and it is thus impossible to fully verify the theorem by numerical methods. Nevertheless, numerical results can provide supporting evidence for the theorem.

For the ferromagnetic Ising models, we fix a one-dimensional lattice of six spins ($L=6$) with periodic boundary conditions, and we determine the coupling constants as follows. We sample the coupling constants $J_{12}, J_{23},\ldots, J_{61}$ from independent exponential distributions of mean 1.0. 
We set $J_{21}, \dots, J_{16}$ by the symmetry $J_{ij}=J_{ji}$ and any other coupling constants to zero. We set the magnetic fields to $H_i = (1+\alpha_i) (J_{i,i+1} + J_{i,i-1})$ with indices modulo 6, where $\alpha_i$ are random numbers sampled from independent exponential distributions of mean 0.5 so that the constants satisfy $H_i \geq \sum_j J_{ij}$. 
We use the transition rate $r_{i}^{\T b}(\Delta\epsilon) = \exp[-\Delta\epsilon/(2\kB\T b)]$. 

For the exclusion process, we fix the number of sites to $L=8$ and sample the number of particles $N$ randomly from 1 to 7. We generate a random monotone potential energy by $e_k = \sum_{\ell=1}^k \alpha_\ell$ for $k=1,\dots,8$, where $\alpha_\ell$'s are sampled from independent exponential distributions of mean 1.0. Since $\alpha_\ell \geq 0$, the potential energy satisfies $e_1 \nobreak\leq \nobreak\cdots\nobreak \leq \nobreak e_L$. We set the transition rates to $r^{\T b}_{k+1\vert k} = \gamma_k \exp[(e_k - e_{k+1}) / (2\kB\T b)]$ and $r^{\T b}_{k\vert k+1} = \gamma_k \exp[(e_{k+1} - e_k) / (2\kB\T b)]$ for $k=1,\dots,7$, where $\gamma_k$'s are sampled from independent exponential distributions of mean 1.0. 

\begin{figure}
\includegraphics[width = 1\columnwidth]{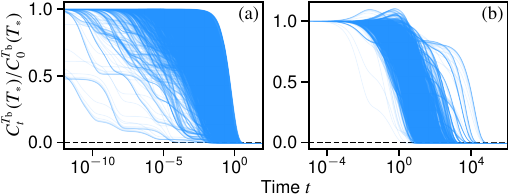}
\caption{Numerical test of the no-Mpemba theorem over models with random parameters. The time evolution of $C_t^{\T b}(T_\largestar) / C_0^{\T b}(T_\largestar)$ is plotted for $\kB \T b =1$ and $\kB T_\largestar \in \{10^{k/3} \mid k=0,1,\dots,12\}$. (a) Ferromagnetic Ising models in the strong magnetic field regime. We generate 500 samples of the random parameters. (b) One-dimensional exclusion processes with monotone potentials. We generate 500 samples.
\label{fig:no-Mpemba}}
\end{figure}

We numerically solve the time evolution of these random models and plot $C_t^{\T b}(T_\largestar) / C_0^{\T b}(T_\largestar)$ for various values of $T_\largestar$ in Fig.~\ref{fig:no-Mpemba}. All curves of $C_t^{\T b}(T_\largestar)$ are above the horizontal line at zero up to a numerical error of $10^{-10}$. This result provides evidence of the no-Mpemba theorem for ferromagnetic Ising models and exclusion processes.

\clearpage

\end{document}